\documentclass{iopart}
\eqnobysec

\usepackage{graphicx}

\begin{document}

\newcommand{\MEMO}[1]{{\bf #1}}
\newcommand{\LATER}[1]{}
\newcommand{\SAVE}[1]{}
\newcommand{\HIDE}[1]{}
\newcommand{\OMIT}[1]{}
\newcommand{\text}[1]{{\rm #1}} 
\newcommand{\Ham}{{\mathcal H}}
\newcommand{\Struc}{{\mathcal S}}
\newcommand{\HH}{{\bf H}}
\renewcommand{\ss}{{\bf s}}
\newcommand{\nn}{{\bf n}}
\newcommand{\rr}{{\bf r}}
\newcommand{\half}{{\frac{1}{2}}}
\newcommand{\RR}{{\bf R}}
\newcommand{\la}{\langle}
\newcommand{\ra}{\rangle}
\newcommand{\uu}{{\bf u}}
\newcommand{\uhat}{{\hat{\uu}}}
\newcommand{\KK}{{\bf K}}
\newcommand{\kk}{{\bf k}}
\newcommand{\khat}{{\hat{\kk}}}
\newcommand{\qq}{{\bf q}}
\newcommand{\QQ}{{\bf Q}}
\renewcommand{\AA}{{\bf A}}
\newcommand{\BB}{{\bf B}}
\newcommand{\cc}{{\bf c}}
\newcommand{\FF}{{\bf F}}
\newcommand{\EE}{{\bf E}}
\newcommand{\MM}{{\bf M}}
\newcommand{\PP}{{\bf P}}
\newcommand{\JJ}{{\bf J}}
\newcommand{\JJmono}{{\JJ_{\rm mono}}}
\newcommand{\mudrift}{{\mu_{\rm drift}}}
\newcommand{\PPtot}{{\PP_{\rm tot}}}
\newcommand{\MMtot}{{\MM_{\rm tot}}}
\newcommand{\tBB}{{\tilde{\BB}}}
\newcommand{\tEE}{{\tilde{\EE}}}
\newcommand{\tPP}{{\tilde{\PP}}}
\newcommand{\tPhi}{{\tilde{\Phi}}}
\newcommand{\tDelta}{{\tilde{\Delta}}}
\newcommand{\hatGG}{{\widehat{G^2}}}
\newcommand{\TcMF}{{T_c^{\rm MF}}}
\newcommand{\Ftot}{{F_{\rm tot}}}
\newcommand{\LL}{{\bf L}}
\newcommand{\LC}{{\mathcal{L}}}
\newcommand{\BC}{{\mathcal{B}}}
\newcommand{\Ncal}{{\mathcal{N}}}
\newcommand{\sent}{{s}}   
\newcommand{\beq}{\begin{equation}}
\newcommand{\eeq}{\end{equation}}
\newcommand{\eqr}[1]{(\ref{#1})}
\newcommand{\Jspin}{{J_{\rm spin}}}
\newcommand{\tpair}{{t_{\rm pair}}}
\newcommand{\tring}{{t_{\rm ring}}}
\newcommand{\rot}{{\rm rot}}
\newcommand{\tot}{{\rm tot}}
\newcommand{\Kstiff}{{K}}    



\title{The ``Coulomb phase'' in frustrated systems}

\author{Christopher L. Henley
\address{Dept. of Physics, Cornell University}}


\begin{abstract}
The ``Coulomb phase'' is an emergent state for
lattice models (particularly highly frustrated
antiferromagnets) which have local constraints that can be
mapped to a divergence-free ``flux''.  The coarse-grained
version of this flux or polarization behave analogously
to electric or magnetic fields; in particular, defects
at which the local constraint is violated behave as
effective charges with Coulomb interactions. 
I survey the derivation of the characteristic power-law
correlation functions and the pinch-points in reciprocal
space plots of diffuse scattering, as well as applications
to magnetic relaxation, quantum-mechanical generalizations,
phase transitions to long-range-ordered states, and the
effects of disorder.
\end{abstract}

\maketitle

\section{Introduction}
\label{sec:intro}

\subsection{The basic idea}

A class of interesting lattice systems in solid state physics and 
statistical mechanics have ground states which (to a first approximation)
are highly constrained yet highly degenerate -- that is,
more or less, the current definition~\cite{ramirez-review}
of a ``highly frustrated'' magnet.  
A fundamental problem in handling such systems theoretically
is that of ``navigation'' among these states.  Is there a way to label
and enumerate the states so as to carry out a sum over them, and
evaluate a partition function?    Or if a particular state is 
somehow to be ``selected'' out of the ensemble,~\footnote{
For example, the classical state which optimizes the energy gain 
from some perturbation term in a classical Hamiltonian, 
or from quantum fluctuations in  a quantum Hamiltonian.}
 how do we find that ``needle in a haystack''?

One of the standard answers is a coarse-graining, that discards most 
of the information in the configurations and keeps local averages of
certain quantities, thereby converting the lattice problem into
a continuum model.  For this to be fruitful (and physically
meaningful) the quantity being averaged ought to be conserved
(It might also be the order parameter of some long range order,
but our models are liquid-like in the first approximation --
that is just paraphrasing the characterization in the first sentence.)
In ordinary (off lattice) liquids, this is rather mundane:
local particle densities and momentum densities.  But in the
lattice models I have in mind, we are back in the wilderness:
the microscopic degrees of freedom
(e.g. spins) usually don't have a conservation law, and 
(it being a lattice model) there is no momentum conservation.

Fortunately, in quite a few cases, another kind of conservation
is hidden in the first-order ensemble: a constraint on the 
total spin (or other degree of freedom) surrounding each 
lattice point.  That allows us to map each microstate of 
local variables into a
configuration of (weighted) arrows living on the bonds of the lattice,
such that the signed sum of the arrow weights into every vertex 
(outwards minus inwards) is exactly
zero in any allowed configuration.  Such arrows are called
``lattice fluxes'' since this is exactly the zero-divergence
condition an electric or magnetic flux would satisfy (in the
absence of sources), if the field were constrained to lie 
along those lattice edges and its flux could only take
discrete values.

The desired emergent vector field $\PP(\rr)$
is the coarse graining of this 
flux:~\cite{youngblood,huse-dimer,henley-pyrice,isakov}
that is, the mean value of
the lattice fluxes over a volume centered at $\rr$ 
much bigger than a lattice consstant, but much smaller
than the system size; the divergence condition 
becomes $\nabla \cdot \PP(\rr)=0$.  
Furthermore it turns out [Sec.~\ref{sec:coarse-grain}]
that the effective coarse-grained free energy 
has the form $\int d^d \rr \frac{K}{2} |\PP|^2$, 
exactly the form of the field energy of an electric
(or magnetic) field.

There are many ways to apply
this analogy to find the long-distance behaviors
of our constrained model.
Since the probabilities according to this free energy are
Gaussian, one can compute 
practically any desired expectation. In particular, 
the ``spin'' correlations depend on separation $R$
with the functional form of a dipole-dipole interaction,
proportional to $1/R^d$ in $d$ dimensions.
It was somewhat surprising to find such a slow decay
(with a divergent correlation length) in such a liquid-like,
maximally random system.  Such states (in three dimensions)
have acquired the name ``Coulomb phase''.

The conditions for a Coulomb phase are that
\begin{itemize}
\item[(C1)] 
each variable can be mapped to a signed flux $\PP_i$ running along bond $i$;
\item[(C2)] 
the variables obey hard constraints, such that 
the sum of the (incoming) fluxes at each (parent) vertex is zero;
\item[(C3)] 
the system is in a highly disordered phase, without any long range ordered
pattern.
This may be called ``liquid-like'' to express the disorder coupled with 
strong local correlations  [implicit in (C2)].
\end{itemize}

\subsection{Rearrangements}
\label{sec:rearrange}

What local rearrangements are permitted by the flux constraint?  
If I flip the variable on
a bond adjacent to parent lattice site $\alpha$, so as to change
its flux from incoming to outgoing,
I must flip the flux in the opposite way on one of the other bonds.
Thus, the natural rearrangements (either in a simulation, or a real
system) are entire {\it loops}, which are sometimes called ``Dirac strings''.
As one traverses the string in a particular direction. 
the sense of the flux arrows is always the same with respect to the
walking direction; the local rearrangment reverses this sense, 
from  always forwards to always backwards or vice versa.
Thus, if the string extends across the whole system, 
the rearrangement changes $\PP$, but if the string closes, then 
$\PP$ is unchanged.   
[An example of such an update is shown 
later in Fig.~\ref{fig:loops-all}(a).]

\subsection{Outline of the article}

The aim of this review is to survey the lattices and models
(and real materials) in which a Coulomb phase are found
(Sec.~\ref{sec:examples});
to walk 
through the derivation of its power-law correlations
(Sec.~\ref{sec:coarse-grain});
and to highlight some interesting ways that Coulomb-phase 
ideas have been deployed to solve problems in frustrated systems. 
These include,  topological defects that (in the case of 
dipolar spin ice) are (emergent) {\it magnetic monopoles} 
(Sec.~\ref{sec:charges}); dynamics (Sec.~\ref{sec:dynamics});
quantum-mechanical generalizations (Sec.~\ref{sec:quantum});
transitions to ordered phases (Sec.~\ref{sec:transitions});
and quenched disorder (Sec.~\ref{sec:disorder}).

Although this review touches on many aspects of
frustrated spin models (gauge theories, disorder,
quantum dimer models, dynamics), it does not aim
or claim to review any of these major topics,
except for particular instances that happen to 
be tractable using the Coulomb-phase notions.
Indeed, even concerning the Coulomb-phase
developments to date, I have not tried to 
exhaustively survey all papers --  I only try
to represent each aspect of the topic in the sections.
My choice of particular results to highlight
is driven by the motives of presentation,
and does not always imply priority or importance.

\OMIT{and point out some open questions in this topic.}

\section{Examples}
\label{sec:examples}

How can we realize a Coulomb phase?
This breaks up into two questions:
which {\it models}, on which lattices, have 
the requisite constraints? (Sec.~\ref{sec:lattices}).
And, which {\it physical} systems realize such models
(the rest of this section).  Of course, though two physical
systems may realize mathematically equivalent models,
quite likely different kinds of quantities are 
experimentally accessible in the respective systems.

I will also point out that one might view the 
constrained system as either  a $T=0$ or $T=\infty$ 
limit of some statistical mechanics model.
If we consider an enlarged configuration
space in which the constraint is not forced, but its violation
costs energy (as I do in Sec.~\ref{sec:charges}), the
constrained model is the $T=0$ limit.
On the other hand, within the constrained ensemble we might
consider adding a Hamiltonian which breaks the degeneracy
among states -- examples are kagom\'e ice (Sec.~\ref{sec:kag-ice})
and the perturbations driving transitions in Sec.~\ref{sec:transitions} --
in which case the basic ensemble is the $T=\infty$ limit.

\subsection{Lattices}
\label{sec:lattices}

\SAVE{These lattices are typical for 
highly frustrated magnets~\cite{ramirez-review}.}

What models give rise to Coulomb phases?
First I will talk about the lattices, and then the degrees of freedom.

\subsubsection{Parent and medial lattices}

First of all, we always have a ``parent'' lattice $\BC$
which is bipartite, i.e. one that can be 
partitioned into (equivalent) sublattices of even and odd nodes
such that every bond connects an even node to an odd node.
Degrees of freedom -- call them ``variables'' for short --
live on these bonds, i.e. on sites of ``medial'' lattice  $\LC$
consisting of the bond midpoints.
(I will call sites of the parent lattice ``nodes'' simply to
help keep them straight from ``sites'' of the medial lattice.)

Conversely, I call $\BC$ the ``premedial'' lattice of $\LC$.
(In the literature, $\BC$ is often sloppily called the ``dual'' 
lattice, which properly means something different.  
For example, the premedial lattice
of the kagom\'e lattice is the honeycomb, but the dual of
the kagom\'e is the dice lattice.)

See Table~\ref{tab:lattices}.
The most important three-dimensional example is the pyrochlore
lattice (diamond as parent lattice).  
The simplest realization is the ``$B$'' sublattice 
(octahedral sites) of the
spinel (usually oxide) structure. The other important realization
is the pyrochlore crystal structure.  
(That is a large family of oxides, which
contain {\it two} interpenetrating pyrochlore lattices, 
each occupied by a different species of cation.)

An example which deserves
more attention is the ``half-garnet'' lattice.
The magnetic lattice in a garnet
consists of two interpenetrating copies of this lattice.
\footnote{
The ``hyperkagom\'e'' lattice~\cite{Ok07,Ho07}
is equivalent to the half-garnet lattice,
so long as only first neighbor bonds are taken into account,
but has less symmetry: one primitive cell of hyperkagom\'e
is two primitive cells of half-garnet.}
The parent lattice is the ``Laves graph of degree three'',
in which each vertex has three bonds forming 120$^\circ$ angles,
and its Bravais lattice is bcc; thus, its 
symmetry is just as high as the pyrochlore's.

I also included the ``octahedral'' lattice, which gets reinvented 
from time to time as a toy model because its parent lattice is 
simple cubic, suitable for simple-minded theorists.
Finally, the ``sandwich'' lattice consists of two kagom\'e layers linked 
by an additional triangular layer, and
models the antiferromagnet SrCr$_{8-x}$Ga$_{4+x}$O$_{19}$), so it is
essentially two dimensional.
There are experimental realizations for most of these lattices.

\begin{table}
\caption{Parent lattices and medial lattices.
The table shows the spatial dimension $d$, the
parent lattice with 
its coordination $Z_\BC$,
and Bravais lattice, 
the medial lattice with its coordination $Z_\LC$, and the number
of sites per cell.
The length $\ell_{\rm loop}$
of the shortest loop (same in either lattice) is also given.
Finally, a reference for the lattice is given.
The clusters formed around each node of the parent lattice 
have $Z_\BC$ sites (forming triangles, tetrahedra, or
octahedra).}
\begin{tabular}{|rl rl lr rr l|}
\hline 
$d$ & parent      &   $Z_\BC$  & Brav. & medial        & $Z_\LC$ & 
   sites & $\ell_{\rm loop}$ & Ref. \\
         & lattice $\BC$  &  & latt. & lattice $\LC$ &         
   & /cell  &                &     \\
\hline 
2 & square & 4 & sq. &   checkerboard & 6 &    2 &  4 &\cite{moessner98a,moessner98b} \\
  & honeycomb & 3 & tri. &   kagom\'e     & 4  &    3 &  6 &\\
  &  4-8 lattice & 3 & sq.  & ``squagome'' & 4 &   4+2 & 4 &\cite{squagome}\\
  & diamond & 3,4 & tri. &  ``kagom\'e''     & 5,6 &  7 & 6 & \cite{broholm-SCGO} \\
  & bilayer  & &     &   sandwich''     &         &   &   &                     \\
\hline 
3  & simple cubic     &  6 & s.c. &  octahedral    & 8 &    3  & 4 &\cite{chui77,hermele,pickles} \\
   & diamond   & 4 & f.c.c. &   pyrochlore    & 6 & 8 &  6  &\\
   &  Laves graph & 3 & b.c.c. &  half-garnet  & 4 & 4? & 10 & \cite{petrenko} \\
\hline 
\end{tabular}
\label{tab:lattices}
\end{table}

For pedagogical purposes, I will use two-dimensional lattices
in all figures, even though (see Sec.~\ref{sec:height}) these
are {\it not} quite bona fide Coulomb phases. 
The reader should view them somewhat more as analogies 
(to $d=3$) rather than as examples.

\subsubsection{Degrees of freedom}

\begin{figure}[ht]
\includegraphics[width=1.0\linewidth]{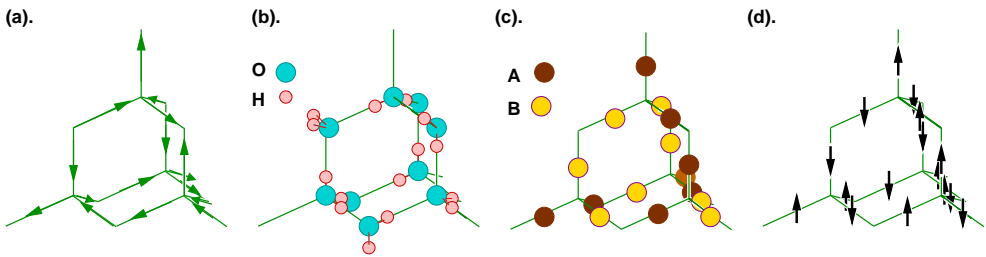}
\caption{
Mappings of the ice model on the diamond lattice
(green edges).
(a). Polarization arrows of the abstract ice model.
(b). Water ice
(c). Compound of species $A$ and $B$
(d). Ising ground state. 
Each configuration shown in (b,c,d) maps to the arrow pattern in (a).}
\label{fig:pyro-examples}
\end{figure}

The most common models reduce to either ice-models or dimer models.
An ice model is defined on a parent lattice with even coordination
number $Z_\BC$ (also called ``six-vertex'' model in the usual
case $Z_\BC=4$.) Every edge carries an arrow, 
and at every vertex the ``ice rule'' constraint is obeyed,
namely half the vertices point in and half point out~\cite{bernal,pauling}.
This arrow obviously {\it is} the flux.  
[See Fig.~\ref{fig:pyro-examples}(a).]

A dimer means an object which covers two nodes of the parent
lattice.  Unless otherwise specified, a dimer covering also
satisfies the condition that every node is covered by {\it exactly}
one dimer: dimers never overlap, and no node is left uncovered.
To define the flux, we draw an arrow of weight $Z_\BC-1$ 
along each dimer-occupied edge, pointing from the even to the odd node;
and arrows of weight $1$ along each unoccupied edges, pointing in the
opposite direction.  (Obviously a different overall normalization of
the flux could be used.)  Illustrations of this weighting may be
found in Fig.~\ref{fig:WL}(a,b,d).

\subsection{Water ices}

I now turn to the different kinds of physical realization, starting
with real ice.
In crystalline H$_2$O, the O atoms form a diamond lattice with hydrogen bonds
to all four neighbors.~\footnote{
The usual form of water ice uses the hexagonal diamond lattice.
The cubic diamond lattice used in the model has very similar behaviors, but
is nicer theoretically owing to its higher symmetry.}
Two of the these adjacent H's are covalently 
bonded to the O  in question (as H$_2$O), while the other two H's belong
to the neighboring O's.  [Fig.~\ref{fig:pyro-examples}(b)].
The variables representing a configuration of
the H's are arrows along the lattice bonds, and
clearly satisfy the ``ice rule''.
This is the first highly frustrated model to be analyzed, 
by Bernal~\cite{bernal} and Pauling~\cite{pauling}
in the 1930s. 

There are many structures (often ferroelectrics) having hydrogen bonds
such that the low energy structures realize some kind of ice rules.

\subsection{Lattice-gas orders}
\label{sec:lattice-gas}

An important way to realize constraints is when some kind of
mutually repelling particle occupies the medial lattice sites.
If we constrain the overall filling $n$ by these particles to 
certain rational fractions, and if the interactions are only
nearest neighbor, it is easy to see the ground states are
those in which {\it every} triangle or tetrahedron has the 
same filling.  If have two
species $A$ and $B$ which attract each other more than their
own kind, with filling $n$ of species $A$ and $1-n$ of species $B$,
as suggested for
CsNiCrF$_6$ \cite{zinkin97,banks}
we get exactly the same ensemble.

When $n=1/2$, we can map the configurations
to an ice model
[Fig.~\ref{fig:pyro-examples}(c)]:
each occupied site becomes an even-to-odd arrow on the corresponding bond 
of the parent lattice; each vacant site becomes a bond in the
opposite direction~\cite{anderson}.
Another special filling is $n=1/Z_\BC$, in which 
case we map the occupied sites to dimers.

One realization is in oxides where the cations 
have a mix of two valence states, corresponding to species
$A$ and $B$; an example of this is magnetite Fe$_3$O$_4$, 
which has an equal mixture of Fe$^{+2}$ and Fe$^{+3}$
on the spinel-B-sites~\cite{anderson}.

\SAVE{Anderson's pioneering paper on the pyrochlore-lattice
antiferromagnet~\cite{anderson} was actually meant to model
inverse spinels, with a mixture of cations.}

\subsubsection{Heavy-electron spinel LiV$_2$O$_4$}

Another example is the metallic spinel
LiV$_2$O$_4$~\cite{fulde-LiV2O4}, 
which exhibits heavy-fermion behaviors
despite the absence of $f$ electrons.
With ion charges Li${+1}$ and O$^{-2}$), charge balance
demands (on the vanadium occupied B sites) 
$n=1/2$ of V$^{+3}$ and the rest of V$^{+4}$.
Fulde and collaborators~\cite{fulde-LiV2O4,fulde}
proposed quantum-mechanical models in which the
correlated electrons can hop with amplitude $t$
on the frustrated lattice.~\footnote{
Since the orbitals are actually $d$ orbitals,
I have glossed over several complications of
orbital degeneracy and orbital dependent
hopping amplitude, as well as spin.}
They work from the limit of strong Coulomb
repulsion, modeled discretely in the spirit of the
Hubbard model; an on-site term effectively 
constrains the occupancy to at most one electron
per site, plus the inter-site repulsion $V$
which is responsible for the flux constraint.

\subsection{Antiferromagnets}
\label{sec:afms}

The remaining kinds of realization are magnetic.

\subsubsection{Relation of antiferromagnetic Hamiltonian to constraint}  

\HIDE{For future use, we define four 
vectors $\uu_m$ pointing along the even-to-odd bond directions
in the diamond lattice.}

Define $\LL_\alpha$ to be the total spin on
the cluster surrounding site $\alpha$ of the
parent lattice:
   \beq
      \LL_\alpha \equiv \sum _{i\in \alpha} \ss_i
   \label{eq:L}
   \eeq
(denoted by ``$i\in\alpha$'').
We want the Hamiltonian to constrain
   \beq
         \LL_\alpha = 0
   \label{eq:Lzero}
   \eeq
for all $\alpha$.
Next, say the Hamiltonian can be written as
  \beq
       \Ham_{\rm spin}
       = \half \Jspin \sum _\alpha \LL_\alpha^2.
  \label{eq:Hspin}
  \eeq
Manifestly, this gives what we wanted:
the classical ground states are
any configuration satisfying \eqr{eq:Lzero}, 
-- the net spin in every cluster is zero.
Furthermore, all such configurations are degenerate.

But in fact, if one simply expands the square in
\eqr{eq:Hspin}, one gets an antiferromagnet
with coupling $J_{ij}=\Jspin$ if 
$i$ and $j$ belong to the same cluster, and
zero otherwise.  In the example lattices 
(all except the octahedral) where the 
every cluster  is a triangle or tetrahedron,
that just means the nearest neighbors: {\it the
ground states of the nearest-neighbor antiferromagnet
are highly degenerate, but every state satisfies the 
constraint \eqr{eq:Lzero} around every $\alpha$.}

The ground state ensemble of a system obeying constraints
and lacking any kind of long range order
may in general be called a ``cooperative
paramagnet'' (sometimes called ``classical spin
liquid'' which means the same thing.)  It bears the same
relation to the usual high temperature paramagnet

Next I explain the three different situations in which 
antiferromagnets can form Coulomb phases~\cite{henley-pyrice}.

\subsubsection{Ising antiferromagnet}

The ground states of the pyrochlore lattice Ising antiferromagnet
have two up and two down spins in every tetrahedron.
The Ising model is not realistic in its own right for
three-dimensional lattices (mainly because their symmetry
is incompatible with the special axis of the Ising spins),
but many other models map to it.
For example, the pyrochlore  Ising antiferromagnet's ground states 
map 1-to-1 onto those of diamond-lattice 
ice model~\cite{anderson,liebmann86}.
Each spin $t_i =+ 1(-1)$ maps to an arrow 
pointing along the corresponding diamond lattice edge, 
in the positive (negative) sense from the even to the odd vertex.
[See Fig.~\ref{fig:pyro-examples}(d).]

Notice that, if we turn on an external field tuned to 
the appropriate size, the ground states
(on any of our lattices) have $Z_\BC-1$ spins up and one down
on every cluster, so this maps to a dimer covering.

\subsubsection{Isotropic Heisenberg model: ``classical spin liquid''}

This state arises when the spins are classical vectors (or 
can be treated as such), 
{\it if}  all states in the continuous manifold satisfying
\eqr{eq:Lzero} are more or less equally likely.  
In this phase,  every vector component of the spins
satisfies the flux constraint;  thus, 
the polarization field has indices not only for directions in
space, but also for the three spin components.

\subsubsection{Isotropic Heisenberg model:``Emergent discrete spins''}

This state can arise in the classical models (mainly the triangle-based lattices)
that manifest ``order by disorder'' in the sense of 
Moessner and Chalker~\cite{moessner98a,moessner98b};
it also arises in {\it all} quantum models with sufficiently large
spin length $S$, when $T$ is low enough that spin-wave energies favoring
collinear or coplanar states are important~\cite{shender,henley-OD,kagome}.
We get a discrete subset of spin states in which 
all spins are collinear along a given axis 
(or coplanar when the clusters are triangles).
The ensemble of this discrete subset is some sort of
Ising model or coloring model, still satisfying 
the constraint \eqr{eq:Lzero}.

\SAVE{It is realistic to model a quantum spin system with large $S$ 
as having well-defined, classical spin directions 
(since the uncertainty bound is $\propto 1/\sqrt{S}$).
However, different classical states have unequal values for 
the total zero-point energy of the harmonic spin-wave 
modes describing quantum fluctuations around those states,
which can be thought of an an additional effective Hamiltonian
term of order $J/S$.}

\SAVE{If we consider the effective
interactions even more carefully, there may be terms
that break the remaining degeneracies and give, in 
principle, a long-range ordered state.  In real 
antiferromagnets, that is likely to happen due to
residual further-neighbor interactions, or to spin-lattice
couplings leading to structural distortions.}

\subsection{Spin ice}
\label{sec:spin-ice}

In ``spin ice'' pyrochlore magnets 
Dy$_2$Ti$_2$O$_7$ and Ho$_2$Ti$_2$O$_7$
local $\la 111 \ra$  
spin anisotropies (additional to \eqr{eq:Hspin}) 
reduce the ground state manifold to 
effective Ising states ~\cite{harris97b,ramirez-spinice},
as reviewed in \cite{spin-ice-review}.
The easy axis is the bond direction $\uhat_i$:
$\ss_i  = t_i \uhat_i$ with $t_i=\pm 1$.  
The actual interactions are {\it ferromagnetic}, $J_F<0$.
Since $\hat  \uu_m \cdot \hat \uu_{m'} = -1/3$ for 
neighboring spins, 
the Hamiltonian in terms of $\{ t_i \}$
reduces to an Ising model with 
{\it antiferromagnetic} $J=- J_F/3$.
In terms of the original model, the ground states
of this ensemble literally implement the ice rules,
i.e. in each tetrahedron (surrounding a parent lattice
node),  two spins point in and two point out.

\subsubsection{Kagome ice}
\label{sec:kag-ice}

When a (not too strong) magnetic field is placed on spin ice 
along (say) the $[111]$ direction, it selects out a subset
(still with extensive entropy) of ground states.
Namely, every spin on a bond of the parent lattice
in the $111$ direction is parallel to the field.
The three remaining sublattices are still free to
fluctuate;   but they form disconnected kagome {\it layers},
so we now have a stack of independent two dimensional systems,
hence this system (realized experimentally)
is called ``kagome ice''~\cite{kag-ice}.
The parent lattice of each layer is a honeycomb lattice; the
spin constraint in the layer is two in/one out on (say) the
even nodes, and oppositely on the odd nodes.  These configurations
map exactly to those of a dimer model on the honeycomb lattice. 
(The dimer positions 
correspond to the single spin with its in/out sense opposite
to the others in its triangle.)

\subsubsection{Dipolar spin ice}
\label{sec:dipolar-spin-ice}

The actual spin ice materials Ho$_2$Ti$_2$O$_7$ and Dy$_2$Ti$_2$O$_7$ 
are well approximated as having nothing but (long-ranged) 
{\it dipolar} spin interactions, rather than
nearest-neighbor ones.  Although this model is 
clearly related to the ``Coulomb phase'', I feel it
is largely an independent paradigm with its own 
concepts that are different from the (entropic) Coulomb
phase that most of this review is about.

\begin{figure}[ht]
\includegraphics[width=0.68\linewidth]{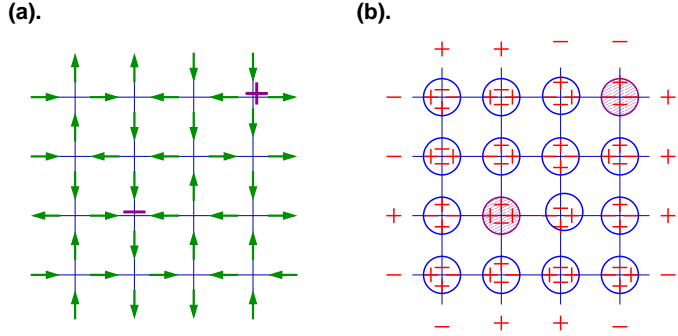}
\caption{Dipolar spin ice and monopoles.
(a). A configuration obeying ice rules, except for 
defects of charge $Q=+2$ and $-2$ (marked).
(b). When each dipole is replaced by a pair of
charges (red $+$ and $-$ symbols) at the ends of its bond, 
these cancel out at every node of the parent lattice
where the ice rule is obeyed (open blue circles), 
but add to a total proportional to $Q$ at defect nodes 
(shaded purple circles).}
\label{fig:monopoles}
\end{figure}

The key fact is that all microstates satisfying the ice
rules have very nearly the same energy~\cite{isakov-whyice};
this was first appreciated numerically~\cite{spinice-dipolar} 
(after the technical problems of simulating long-range interactions
in periodic boxes were sorted out).  The simple explanation
was given in ~\cite{castelnovo-mono}.  It's a good
approximation -- for the far field, anyhow -- to replace
each point dipole by a pair (so-called ``dumbbell'')
of opposite effective charges $\pm q$
having the same dipole moment $\mu$, and we might as well separate 
them by the bond length $d$ of the parent lattice, thus
$\mu = qd$.
So, all the effective charges sit on parent lattice nodes, and
each node gets contributions from four dipoles 
(see Fig.~\ref{fig:monopoles}).  Indeed, so long
as the spin configuration satisfies the ice rules, those four
contributions add to {\it zero} on {\it every} node, and (within
this approximation) the interaction energy is zero (I omitted 
a configuration-independent constant representing the interactions
between the two charges forming each dipole).
In principle, small multipole interactions, representing
the difference between the actual interactions of point dipoles
and those of the effective charges, will break the degeneracy
and lead to ordering at much lower temperatures.

\SAVE{Skipping Nagle dimer models}

\section{Coarse-graining, Fourier mode fluctuations,  
and long-range correlations}
\label{sec:coarse-grain}

\HIDE{NEED TO CITE \cite{henley-pyrice,isakov}}

In this section, I present the steps leading to power-law 
correlations using the framework of a continuum theory, 
in the ideas appear more transparently.

\subsection{Polarization field and effective free energy}

\SAVE{A side remark: if we get flux only modulo $Z_2$, we
get $Z_2$ topological order...}

The polarization field is the key object of the Coulomb phase.
As already laid out, our system has an extensive
entropy of ground states (degenerate at this order).
How to handle this?
Recall the discrete fluxes  $\PP_i$, which we defined
along every bond of the parent lattice.
Define a (whole-system) polarization density $\PP \equiv \sum _i \PP_i/\text{volume}$.
This is a good choice for writing thermodynamic functions,
since the ground-state entropy density evidently depends on it.

Recall the point of Sec.~\ref{sec:rearrange}, that 
rearrangements (within the flux constraints)
must occur along strings  which follow flux arrows. 
Now, if $\PP$ is large, most fluxes have a positive
component in that direction and the typical string crosses the
system -- a closed string would require half of its fluxes to have
a backwards component, but such a local gathering of 
reversed fluxes is unlikely.
On the other hand, when $\PP$ is near zero, there are many closed strings
in any configuration, and hence many rearrangements which preserve $\PP$.
It's clear, then, that the number of configurations $\Ncal(\PP)$ with a given 
polarization is maximum at $\PP=0$ and goes to zero as $\PP$ approaches its 
saturated value.    So let's define an entropy density 
   \beq
       \sent(\PP) \equiv \lim _{V\to \infty} \frac{\ln \Ncal(\PP)}{V} 
   \label{eq:sigma-def}
   \eeq
which is maximum at zero.
\SAVE{ the number of states (and hence ensemble weight) behaves as $\exp(-\sent(\PP) V$.}

Next, so long as the ensemble is a ``liquid'' lacking long-range orders,
the {\it local} polarization (if we divide the system into smaller boxes)
is fluctuating, and not too strongly correlated from box to box.
Then the Central Limit Theorem says the ensemble probability (= number of ground
states for a given $\PP$) in a system of large volume $V$ 
approaches a Gaussian form
  \beq
     \Ncal(\PP) \propto \exp(-|\PP|^2/2 {\sigma_P}^2)
  \eeq
with a variance $\sigma_P^2 = 1/\Kstiff V$ for some $\Kstiff$.  
Comparing with \eqr{eq:sigma-def}, we see
  \beq
     \sent(\PP) \approx \sent_0 - \half \Kstiff |\PP|^2
  \eeq
for small $\PP$.  (All this was a completely standard argument from basic statistical
mechanics; it is the same reason the free energy is proportional to (magnetization)$^2$ in a paramagnet,
or that one assumes analyticity in the Landau free energy functional.)

Next let's consider the spatial fluctuations of $\PP$, which are more interesting 
(and more measurable!) -- than the functional form of $\sent(\PP)$.
This requires defining a spatially varying 
polarization {\it field} $\PP(\rr)$;  it is 
a coarse-graining, i.e. the average of discrete polarizations
over some neighborhood of $\rr$ (one much larger than
the lattice constant but much smaller than the system size.)
We assume $\PP(\rr)$  varies smoothly.
Corresponding to the discrete flux constraint,
$\PP(\rr)$ satisfies  a divergence constraint
   \begin{equation}
       \nabla \cdot \PP(\rr)=0
   \label{eq:divergence}
   \end{equation}
like a magnetic field without monopoles.

The total free energy (arising entirely from entropy) is the
sum of those in the boxes or averaging volumes into which 
we can divide the system, hence
   \begin{equation}
       \Ftot \Big(\{ \PP (\rr) \Big\})/T =  \text{const} + 
  \int d^d\rr \half \Kstiff |\PP(\rr)|^2, 
   \label{eq:Fcoarse}
   \end{equation}

Eqs.~\eqr{eq:Fcoarse} and \eqr{eq:divergence} look, respectively,
like the field energy of a magnetic (or electric) field, 
and its divergence constraint,  in the absence of monopoles (or charges). 
That electrostatic (or magnetostatic) analogy is fruitful, and is
why this state was dubbed ``Coulomb phase.''

In the case of dipolar spin ice, since the fluxes are parallel to real moments,
the polarization is proportional to the real magnetization $\MM$.
   \beq
           \MM= \mu \PP.
   \eeq
Furthermore, if a region has a net polarization, we have
a field energy of form \eqr{eq:Fcoarse},
but now $K\equiv \mu_0 \mu^2$,
where $\mu_0$ is the permeability of free space, in
the limit $T=0$: it is purely energetic rather than entropic.
At $T>0$, the entropic elasticity gives an correction
to the permeability. (In water ice, which has long range
electric dipole interactions, the analogous contribution 
to the dielectric constant has long been known~\cite{nagle-ice}.)
\SAVE{Since the arrows in real ice are associated with 
real dipole moments, this entropic effect appears there 
as a modification of the dielectric constant.}

\subsection{Pseudodipolar correlations and structure factor}
\label{sec:correlations}


The standard way to evaluate correlations is to transform to Fourier space.
Eq.~\eqr{eq:divergence} gives
        \beq
           \qq\cdot  \PP(\qq) =0
        \label{eq:div-Fourier}
        \eeq
so a naive use of equipartition would give 
$\langle P_\mu (-\qq) P_\nu(\qq) \rangle  = (1/\Kstiff) \delta_{\mu\nu}$.
so \eqr{eq:Fcoarse} gives
       \beq
          F_{\rm tot}= \sum _\qq \half \Kstiff |\PP_\perp(\qq)|^2.
        \label{eq:F-Fourier}
       \eeq
where $\PP_\perp(\qq)$ refers to the components of $\PP(\qq)$
satisfying \eqr{eq:div-Fourier}.  The fluctuations are gotten by
first writing the (trivial) result of equipartition for an
unrestricted $\PP(\qq)$, and then projecting to obtain 
the transverse part:
   \begin{equation}
   \Struc_{\mu\nu}(\qq) \equiv \la  P_\mu(-\qq)  P _\nu(\qq')\ra 
    = \delta_{\qq,\qq'}\frac{1}{\Kstiff}\Bigl( \delta _{\mu\nu}  - 
     \frac {q_\mu q_\nu}{|\qq|^2} \Bigr),
   \label{eq:flucts}
   \end{equation}

\subsubsection{Diffraction consequences}

A physical observable $\Phi(\rr)$ usually has a contribution proportional to $\PP$,
but possibly the correspondence is modulated e.g. by alternating
signs, so we write 
   \beq
       \Phi_a(\rr) = (...) + \sum_\mu c_{a\mu}(\rr) P_\mu(\rr).
   \label{eq:Phi-P}
   \eeq
where $\{ c_{a\mu}(\rr) \}$ is a matrix of coefficients, with the
symmetry of the lattice, and indices $ab$ refer to possible i
components of the observable.
It follows that the observable structure factor, measured in diffraction, 
behaves as~\footnote{
If ``(...)'' in Eq.~\eqr{eq:Phi-P} is a function of local fluctuating
variables that has no conservation law, then ``(...)'' in \eqr{eq:Phi-struct}
adds a non-singular diffuse contribution on top of the features of 
$\Struc_{\mu\nu}(\qq)$.}
   \beq
       \Struc^\Phi_{ab}(\qq) \equiv \la \tPhi_a(\qq)\tPhi_b(-\qq)\ra 
    = (...) + \sum _{\mu\nu\QQ}  f_{ab\mu\mu}(\QQ) \Struc_{\mu\nu}(\qq-\QQ).
   \label{eq:Phi-struct}
   \eeq
where $f_{ab\mu\mu}(\QQ)$ is some sort of structure factor derived from
the $c_{a\mu}$'s, and $\QQ$ are reciprocal lattice vectors.

Two important features may be noted.  First, the polarization constraint
says that a certain linear combination of fluxes is exactly zero
(their projection on the inward direction at a node of the parent lattice).
That translates into a strict zero of the appropriate structure factor,
at a certain wavevector.  In the case of an antiferromagnet, or
spin ice, the constraint is that magnetization is exactly zero 
in each cluster.  Hence, we find a very low scattering near the
zone center (the diffuse intensity grows as a high power of $|\qq|$);
also, the intensity has zeroes along every direction that is a 
symmetry axis of all clusters.

\begin{figure}[ht]
\includegraphics[width=0.77\linewidth]{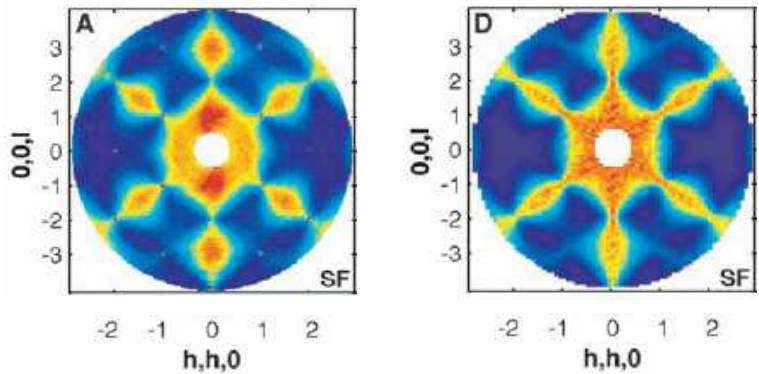}
\caption{Pinch point in diffraction.
From \cite{fennell}.  Spin-flip component of polarized neutron diffuse
scattering from the spin-ice pyrochlore compound Ho$_2$Ti$_2$O$_7$,
in the $(hhl)$ plane of reciprocal space.
Left side, experimental result; right side, Monte Carlo simulation.
The added arrows highlight the ``pinch point'' features at the
wavevectors labeled.}
\label{fig:pinchpoints}
\end{figure}

The second and more striking feature comes
from the second term in \eqr{eq:flucts}.
Although not divergent, it is singular: the ratio has a different
limit at $\qq=0$, depending on the direction of approach.
In reciprocal space, this has the characteristic shape of 
a ``pinch point'', at which the contours of equal intensity
have a roughly triangular shape.
The form factor in \eqr{eq:Phi-struct} translates these to
reciprocal lattice vectors $\QQ$ other than zero, as seen
in date from \cite{fennell}, shown in Fig.~\ref{fig:pinchpoints}.  
(Earlier experiments~\cite{Bra01b} gave less clearcut images:
polarized neutron diffraction is needed to separate the contributions 
from different spin components.)

\HIDE{\cite{moessner98b}, 
gave a heuristic argument for a scaling form in $q_x/q_\perp$
to explain the ``bowtie'' diffraction already observed in simulations.}

\SAVE{See Fig. 1 of \cite{Bra01b}
showing the structure factor in Ho$_2$Ti$_2$O$_7$.}

I discussed in \cite{henley-pyrice} the way that other kinds of
hard constraint also produce sharp features in reciprocal space. 
For example, if you constrain a lattice gas on an fcc lattice
so that every particle has an equal number of occupied
neighbors, you get rings~\cite{deridder}. 
(Similar features are observed in quasicrystals and 
ascribed to local tiling constraints~\cite{henley-RT}).
This constraint superficially resembles that of 
a coulomb phase  as laid out in Sec.~\ref{sec:lattice-gas};
the outcome is different because the fcc lattice isn't the
medial lattice of a bipartite lattice.)

\subsubsection{Real space correlations}
\label{sec:real-space-corr}

Fourier transforming \eqr{eq:flucts} back to direct space gives
   \begin{equation}
       \langle P_\mu (0) P_\nu(\rr)\rangle \cong 
       \frac {c_d}{\Kstiff r^d} 
          \Bigl(\delta _{\mu\nu} -d \hat r_\mu\hat r_\nu \Bigr)
   \label{eq:ice-dipolar}
   \end{equation}
at large separations $\rr$ in $d$ dimensions, 
where $\hat \rr \equiv \rr/|\rr|$ and $c_3=4\pi$.
So the correlations,
which one naively expected to be exponentially decaying
in this ``liquid-like'' state, 
are instead power-law-decaying i.e. {\it critical}-like, 
They have the spatial dependence of a {\it dipole-dipole interaction}.

This  criticality was appreciated in the ice model as early 
as 1973, being detected originally in a simulation~\cite{stillinger}.
The universal explanation (above) how dipolar correlations 
arise from \eqr{eq:Fcoarse} with the divergence condition,
was first put forward in 1981 to explain experiments on
two-dimensional ice-like systems~\cite{youngblood}.
It was noticed in the context of antiferromagnets 
in \cite{henley-BAPS}
and (most influentially) announced as a general idea
(initially for dimer coverings) by Huse {\it et al}~\cite{huse-dimer}.
\SAVE{
\cite{hermele} obtained the $1/|\rr|^3$ behavior for the pyrochlore lattice. }

\SAVE{Ref.~\cite{moessner98b}, Sec.~II~D 1, recognized
that the ground-state constraint in the Heisenberg
pyrochlore antiferromagnet \eqr{eq:Lzero} entails
long-range correlations, but in the absence of
the polarization concept, the argument's form is diffuse,
and it was not possible to predict an explicit functional form.}

We can apply these ideas to the 
nonlocal spin susceptibility near a defect.
Consider a vector-spin antiferromagnet
on one of the lattices in Table~\ref{tab:lattices}, where
the spins are diluted (this corresponds to bond dilution on
the parent lattice).  At a dangling node, one that has just one neighbor,
the flux constraint cannot possibly be satisfied.
Such ``orphan spins'' respond to external fields like free moments,
and in line with Eqs.~\eqr{eq:ice-dipolar} [or \eqr{eq:coulombV}, below]
their perturbation of the surrounding spins decays
as a power law (with oscillations depending on how spins
map to polarizations). This can be modeled within a
Coulomb-phase framework,~\cite{damle}
and used to explain NMR observations
in SrCr$_{9p}$Ga$_{12-9p}$O$_{19}$.

\subsubsection{Calculation for specific lattices}

So far we only obtained the asymptotic functional form near
``pinch points'':  more practically, one would like to
model (approximately) the diffuse scattering over the
entire Brillouin zone for a specific model.
For the spin models of Sec.~\ref{sec:afms}, a systematic way is to allow
$n\to\infty$ where $n$ is the number of spin components.
This ``large-$n$'' approach  is a standard
trick of statistical mechanics, since (in that limit)
the  constraint of unit length becomes irrelevant and
we have a linear problem~\cite{isakov,garanin-canals,canals-garanin}.
This turns out to be an good approximation even for $n=1$.

An ad-hoc ``maximum-likelihood'' approach,~\cite{henley-pyrice},
which can be applied to {\it any} model of our class, is to consider 
the fluxes $P_i$ (representing the model variables)
to be {\it real} numbers, subject to two kinds of constraint:
(i) that $P_i$ take certain discrete values (ii) the usual divergence
constraint.   We then replace (i) by a weighting $\exp(-\half P_i^2/\sigma^2)$,
where $\sigma^2$ is chosen to give the correct variance when the $\{ P_i \}$
are chosen at random (and unconstrained by the divergence condition).
The cluster-variational approach~\cite{yoshida} is related.
\LATER{CHECK THIS OUT}

Finally, Villain predicted correlations in ice using
a clever random-walk approximation to a series expansion~\cite{villain},
which has not yet been reconsidered in the literature.
I would suspect this amounts to a Bethe-lattice approximation
(i.e. neglecting the existence of loops in the lattice).

\subsection{Height models: two-dimensional Coulomb phases}
\label{sec:height}

It is even easier to realize a flux constraint in $d=2$ than in $d=3$,
and such models exhibit all the phenomena mentioned here.  However,
they exhibit additional, even more striking, behaviors which are peculiar
to two dimensions, and for that reason they fall outside the scope of
this article.  Here I will just summarize the differences; 
such ``height models'' call for a review paper 
of their own~\cite{BCSOS,bloete,kondev,kondev-4color,raghavan,zeng}.

To see what's different, recall that a divergence-free
field such as our $\PP(\rr)$ can always be written  as
a curl of a vector potential, $\PP (\rr)=\nabla \times \AA(\rr)$.
In three dimensions, $\AA$ is ill defined due to its gauge freedom;
but in two dimensions, $A(\rr)$ has only one component (normal to
the plane) and is an ordinary potential, uniquely defined modulo
an additive constant.  [To see this, note that $\PP_\rot\equiv(-P_y,P_x)$
carries the same information as $\PP$ but satisfies $\nabla \times \PP_\rot=0$,
hence $\PP_\rot = \nabla A$ defines a potential.]

We can think of this as mapping a configuration of the variables
to one of a crystal interface~\cite{BCSOS,bloete}
with profile $z=h(x,y)\equiv A(x,y)$.
The ``Coulomb phase'' corresponds to the ``rough'' phase of 
such an interface model with the well-known form
$F_\tot = \int d^2 \rr \half K |\nabla h(\rr)|^2$.
However, because $h(\rr)$ is uniquely defined, a generic physical variable
(written in terms of field variables)
is not only a linear combination of $\nabla h$ terms, but also has
periodic terms of form $\cos(2\pi m h/ a_\perp)$.  [The repeat offset 
$a_\perp$ depends on the model and the definition of  $h(\rr)$.]  
It can be shown that:
\begin{itemize}
\item[(i)] 
This leads, in the rough phase, to
critical correlations with a parameter (and temperature) 
dependent exponent $\propto T/K$.
\item[(ii)] 
An un-roughening transition occurs
when $T/K$ decreases past a universal ratio, 
in which the system {\it locks} to a particular value of 
$A$, corresponding to {\it long-range order} of the model variables;
\item[(iii)] 
Thermal excited ``defect charges'' (see Sec.~\ref{sec:charges})
do not necessarily destroy the critical phase; rather, they stay
bound in pairs for $T/K$ less than a universal critical ratio;
the unbinding is the ``Kosterlitz-Thouless'' transition familiar
in the XY model.
\end{itemize}

The critical correlations (i) imply singularities in reciprocal 
space, generally displaced at different positions in the Brillouin zone 
than the pinch points.  
Such singularities (in the height field) were called ``zone-boundary
singularities'' in Refs.~\cite{kondev-4color,raghavan,zeng} or (in the
structure factor) ``pi-ons'' in Ref.~\cite{moessner-kagice}.

\SAVE{For most natural models, we are already on the unbound side
of the K-T transition.}

\section{Pseudo-charge defects in Coulomb phases}
\label{sec:charges}

If we change our rules to allow configurations that violate 
the flux constraint, merely including a Hamiltonian which strongly
penalizes the violations so they are
dilute at low temperatures -- what do we get?
\SAVE{(Or small, in continuous-spin models.}
Such a defect may be labeled by its ``charge'' $Q$, equal to the net
(nonzero) flux in the outward sense (at a parent-lattice vertex).
This is a ``charge'' in the sense of Gauss's law, as 
the net flux through any surrounding surface must equal $Q$.
That means the ``charge'' is conserved in time: the net charge in some volume 
can't be changed except by moving it across the boundary, and 
defects can only be created in pairs of opposite charge.
Their detectability from distant measurements, as well as the
conservation properties, are characteristic of 
{\it topological defects}. 

More interesting, in a Coulomb phase, the effective potential between 
defects must have the form of Coulomb's law:
  \beq
      F_{\rm int}(\rr_1,\rr_2)/T = \frac { K Q_1 Q_2} {4\pi |\rr_1-\rr_2|}
  \label{eq:coulombV}
  \eeq 
(in $d=3$).  
The interaction is defined by integrating the partition function, conditional
that defect charges $Q_1$  and $Q_2$ are placed at $\rr_1$ and $\rr_2$, 
and by requiring the result to be proportional to $\exp (- F_{\rm int}/T)$.
That is essentially how one obtains the Coulomb potential from the field energy
in electro- or magnetostatics,  and we must get the same
result here since the field free energy has exactly the same form.
In $d=3$ (but not necessarily in $d=2$), our ``charges'' are {\it deconfined} 
i.e. a defect/anti-defect pair (in a large system) will separate and
have independent positions.  By contrast, if these same defects are
present in a phase with long-range order of the variables, pulling apart 
the defects now carries a cost proportional to $|\rr_1-\rr_2|$: 
in that case, defects are ``confined'' the same way that quarks are.

Some authors have emphasized the notion of ``Dirac string''
meaning the trail of fluxes that got flipped as you pulled
apart a defect/antidefect pair.  It should be realized that
this is a nebulous and not very helpful notion when applied
in the Coulomb phase proper (with smallish polarization), 
since the string's path is not well defined. That is, there 
are many different, equally good ways to represent a defect pair 
by (i) taking some configuration from the undefected ensemble 
and (ii) flipping the fluxes along a string.  It is only in an 
ordered phase (see preceding paragraph), or near the limit of maximum polarization, 
that the Dirac string has a clear meaning.

When the model is a lattice gas of neighbor-repelling particles,
that map to a dimer covering on the parent lattice, a simple 
way to create a defect/antidefect pair is to remove one
particle or dimer (the total flux at the parent sites that the
dimer spanned gets increased/decreased by the flux associated
with a dimer).  Evidently each of the two defects is carrying 
an effective particle number of 1/2, making this an elementary
example of ``fractionalization''~\cite{moessner-tri-RVB}.


In actual water ice, OH$^-$and H$_3$O$^+$ ions (if the hydrogen bond
network is unbroken) are topological defects of charge $Q=\pm 2$ like
the ones we were describing -- except they aren't very mobile, 
since proton transfer is slow at icy temperatures.  
A distinct kind of defect, the Bjerrum defect, is formed
when there are no protons, or two, along a given bond
(maps to a nonmagnetic or doubly magnetic impurity in a spin system).
The Bjerrum defects have charge  $Q=\pm 1$ and are mobile.

\subsection{Thermal consequences}

If the creation energy is $E_Q$ for a defect of charge $Q$, 
the defect density will behave as $n_Q \propto \exp(-E_Q/T)$.  
The specific heat is affected proportionately.
[As in semiconductors, the exponent is $E_Q/T$ even though 
each defect pair costs $2 E_Q$, due to entropy ,
as the defect locations are independent.]

The defect ``charges'' are in the same situation
as ions in a plasma (or in water), or as carriers
in a compensated semiconductor:  we get Debye screening of
the fields.  In consequence, the ``Coulomb''
effective interaction, \eqr{eq:coulombV}, acquires a decaying
factor $\exp(-\kappa|\rr_1-\rr_2|)$, where $1/\kappa$ is the 
Debye screening length, given by
    \beq
            \kappa = \sqrt{ \frac{n_Q K Q^2}{T} }.
    \label{eq:Debye-kappa}
    \eeq
The pseudodipolar polarization correlations [Eq.~\eqr{eq:ice-dipolar} 
acquire the same exponential screening factor.  In Fourier space,
this corresponds to replacing $|\qq|^2 \to |\qq|^2 + \kappa^2$
in the denominator of Eq.~\eqr{eq:flucts}.

How thermal excitation plays out in real systems depends on 
details of the model.  In ``spin ice'', the spin configurations 
are not inherently constrained to obey the constraint,
so at $T>0$ there is indeed a thermal concentration of defects.
The cost 
     \beq
          E_Q= J Q^2
     \eeq
is $2J $ for the minimum charge $Q=\pm 2$.
Following \eqr{eq:Debye-kappa}, then, we expect a 
$\kappa \propto \exp (-J/T)$, and this agrees with the
experimental fit to the structure factor~\cite{fennell}.
Around the reciprocal-space point where the defect-free 
diffuse scattering has a zero and flat basin,
due to the exact cancellation in each tetrahedron, 
we expect an increase of diffuse scattering $\propto n_Q \propto \exp(-2J/T)$
as is also observed~\cite{fennell}.

On the other hand, say one has a dimer model on a triangular
lattice and that dimer constraint is absolute.  It still is
not a Coulomb phase~\cite{moessner-tri-RVB} (since the parent lattice isn't bipartite).
However if a very strong external strain field biases the dimer orientations 
such that one orientation is excluded, the remaining (rhombic)
lattice {\it is} bipartite and develops a Coulomb phase.
(Exactly the same thing  happens with dimers on the fcc lattice.

The experimental example of ``kagom\'e ice'', where  the 
external bias is the magnetic field, is similar: in that case,
the bias causes a two-dimensional Coulomb phase to emerge
from a three-dimensional one.)

Now, if the bias is reduced (but still strong) so that an occasional 
``wrong'' orientation appears by thermal activation, that
merely creates a {\it dipole} in the rhombic lattice's 
polarization field.  As we also know from electrostatics,
an ensemble of dipoles changes the dielectric constant
(or its analog $1/K$, in our case) but does not give Debye
screening, so we continue to have power law correlations.

\subsection{``Magnetic monopoles'' in dipolar spin ice}

In the case of {\it dipolar} spin ice
(recall Sec.~\ref{sec:dipolar-spin-ice}),
what happens at a node where the ice rules
are violated, so the net flux there is $Q$?
The same construction outlined in Sec.~\ref{sec:dipolar-spin-ice},
whereby point dipoles get
replaced by pairs of effective charges $\pm q$, 
must put a net (magnetic) charge of $Q$ on every such defect node,
just as it puts zero charge on every other node.
(See Fig.~\ref{fig:monopoles}).
The total energy of such a state is now given by
the Coulomb interactions of these defect charges:
they are {\it emergent magnetic monopoles}~\cite{castelnovo-mono}.

\OMIT{Perhaps the fact that the defect is a monopole
was already evident from the field force \eqr{eq:field-force}.}

We should be clear in what sense this is a monopole.  
Microscopically, of course,
the laws of nature continue to rule out monopoles. Thus, near to any
defect site, in between the bond directions where there is an excess 
of (say) outward flux, there must be other directions where there 
is a ``counterflow'' of inward flux.  This seems rather analogous to the
current associated with the Bogoliubov  quasiparticle in a 
BCS superconductor.
Its motion across the sample indeed causes a charge transport, proportional to
the difference between the fractions of electron and hole making up the
quasiparticle.  Yet, while traveling in the bulk, the quasiparticle 
actually can have no charge~\cite{bogo-charge}
 (in consequence, there, of the Meissner effect).

Notice that although the ``charge'' interactions in dipolar spin ice 
look formally like those in entropic Coulomb phases, 
the physics is quite different. 
In the entropic case, 
the {\it energy} cost is only the core costs of making the
defects; the effective Coulomb interaction is wholly entropic,
and emerges only as we allow the system in equilibrium to
explore all those many microstates.
By contrast, in the dipolar case, {\it every}
one of the (many) microstates having defects of $+Q$ and $-Q$ at
$\rr_1$ and $\rr_2$ has very nearly the same spin interaction
energy, dependent only on $|\rr_1-\rr_2|$.  

A way to note the difference is that for an ice model in $d=2$,
defect charges or monopoles separated by $R$ feel a $\ln R$ potential 
in the entropic case (just consider the Gauss's law for the model
flux,  confined to the plane) whereas in the dipolar case they
have the usual $1/R$ interaction characteristic of three dimensions.
Another situation that sharply illustrates the difference is
``kagome ice'' (see Sec.~\ref{sec:kag-ice}).
So long as the ice rules are satisfied, that system breaks up 
into decoupled two-dimensional layers.  
But when charge defects are introduced in the dipolar system,
their interaction is isotropic -- it is the same whether the
two defects are in the same or in different layers
(even though each defect is still free only to move within a layer).
By contrast, in the kagome ice phase of model spin ice with
nearest-neighbor couplings, an interlayer defect interaction 
exists at all only to the extent that spins linking the layers
have some fluctuations and mediate it.

One might reckon that the short-range models, in which (entropic) 
Coulomb behavior emerges from a collective state, is less trivial
than the dipolar model, in which we have a bare magnetic energy.
After all, our defect monopoles are not terribly different from the 
emergent monopoles  found at the end of a long thin bar magnet, 
as we were taught in introductory magnetostatics.  
The important difference, and the reason for all
the excitement is that -- unlike the bar magnet poles --
\begin{itemize}
\item[(i)]
our monopoles move freely in response to forces;
\item[(ii)]
their magnetic charges are quantized
(the value depends on material properties), 
\OMIT{In those regards they are bona fide emergent particles.}
 the defect monopoles are mobile
\item[(iii)]
A variety of cute experiments is possible which depend
on the monopoles.
\end{itemize}

I would reserve the term ``observation'' of a monopole for
an experiment isolating a single one.
One might observe quantized jumps in the
induced current  around a conducting ring embedded in a sample
whenever a monopole passes through it~\cite{castelnovo-mono}.
Alternatively, a monopole just inside the sample's surface
creates a characteristic $1/R$ field outside it, which might
be detectable by a scanning magnetic force microscope.

The experiments so
far are instead thermodynamic and transport measurements,
the interpretation of which depends on monopoles being the
elementary excitations.
Some of the experiments are:
\begin{itemize}
\item
   A phase separation between two paramagnetic phases
-- monopole liquid and monopole gas -- as seen in
kagome ice~\cite{phase-separate}
\item
Magnetization dynamics (see below), whereby 
the time derivative of magnetization maps 
to a monopole current~\cite{jaubert-mono}
\item
Measuring the monopole charge by an analog of the 
Wien effect~\cite{sondhi-wien}: that is, 
how the density of monopoles increases in a magnetic field
which reduces the ``ionization energy'', manifested in
a reduction of the spin autocorrelation time as probed
by muon spin relaxation~\cite{bramwell-mono}.
\item
Thermodynamics...\cite{fennell,bramwell-mono,morris,kadowaki}
\LATER{Say a bit better}
\end{itemize}

It should be observed, that many of the experiments basically probe
the existence of thermally excited pseudo-charge defects as could
be found in any Coulomb phase system; only a few of the experiments
test the special ``monopole'' property of the defects in the spin
ice case (namely their literal magnetostatic interaction).

\subsubsection{Correlation length experiment}

A recent experiment~\cite{morris} on spin ice at high magnetic fields
was interpreted in terms of monopoles.
The polarization is near saturation, so
the only freedom is in dilute line-like excitations, of overall density
proportional $\delta P = P_{\rm max} - P$. 
These  are none other than 
the ``world lines'' of Sec.~\ref{sec:world-lines}
and also are genuine ``Dirac strings'';
if the flux constraint is absolute, 
they never terminate. 
But in the presence of a small density $n$ of defects, each 
world line connects a defect and an antidefect (and all defects
are endpoints), the typical length being
$P/n$.
[The connecting-up is well-defined only to the extent that 
world lines don't touch, which is true if $(\Delta P)^2/n$ 
is sufficiently small.]
\LATER{This should be a footnote.}
The numerical agreement between the diffraction width and simulation
results is indirect evidence for the finiteness of the Dirac
strings, and hence the presence of separated defects.

\section{Dynamics}
\label{sec:dynamics}

We could also make predictions for the dynamic fluctuations and
response of the polarization field.

\subsection{Accessibility of states under updates}

As was pointed out (in Sec.~\ref{sec:rearrange}),
the natural update for a model
with a flux is a loop update. 
[An example of such an update is shown 
later in Fig.~\ref{fig:loops-all}(a).]
However, with luck, one can run a simulation 
considering just {\it minimal updates}, i.e. involving the shortest
loops (one plaquette, in the example of 
Fig.~\ref{fig:loops-all}(a), or
a hexagon in the case that the parent lattice is a honeycomb or
diamond lattice).   

Now, in order to equilibrate a simulation, an update move must 
allow one to access every state with a given global flux, 
from every other state with the same flux, given enough steps.
Minimal updates allow access in two-dimensional dimer models,
but seem not to in the diamond-lattice dimer covering, so that
non-local updates are needed.
One method for this is a generalized ``worm'' update~\cite{worm-update}
(originally applied just
to accelerate the dynamics in models for which minimal updates 
allow access).
One creates a defect pair, allows the defects to diffuse till they 
re-annihilate, and then accepts or rejects the final configuration 
according to the usual Metropolis fashion.  

This is the typical mechanism for updating in a real system, too.
That is, even rather dilute thermally excited defects are
essential ``lubricants'' facilitating relaxation. 

\subsection{Dynamics of defects}
\label{sec:charges-dynamics}

Topological defects, even when very dilute, may be crucial in the dynamic 
behavior for two reasons.  First, the inherent dynamics of the Coulomb
phase is presumably local, but a finite update may be insufficient to 
access all states in our ensemble. After all, as I have mentioned,
the natural update follows a loop, and -- given that the Coulomb phase 
has power-law correlations -- one can't rule out that the updating loops 
have a power-law distribution with a tail of long loops.  But if a pair
of defects is created, random-walk for a long time, and re-annihilate,
this accomplishes the same effect as a long update loop.  (If the trails
of the defects never crossed -- a big ``if'' -- that loop would just be 
the joining of those trails.)
The defect pair might be created by thermal activation in a realistic
dynamics, or artificially as a recipe to accelerate some Monte Carlo dynamics.

Defects matter for a second reason in cases, e.g. spin ice
or water ice, where the model polarization is a {\it physical polarization}
(magnetization or electric polarization) that couples to external fields:
they control the relaxation of the polarization.
Let's consider spin ice for concreteness,
and let $T>0$, so there is some density $n(T)$ of thermally excited
monopoles.

Whenever a monopole (defect) moves by $\Delta \rr$, it changes the 
system's total polarization by $\Delta \PPtot = Q\Delta \rr$.    
Firstly, this implies the relaxation rate of the
total polarization is the monopole current,
   \beq
        \frac{ d \PPtot}{dt} = \JJmono.
   \eeq
Secondly, it implies the external (real) magnetic field 
$\BB$ applies a force
    \beq
         \FF_{\BB} = Q \BB
    \label{eq:field-force}
    \eeq
on each monopole defect.  So assuming a linear response
with a mobility coefficient $\mudrift$, 
each defect has a drift velocity $\mudrift Q \BB$.
Finally, the system's total magnetization is $\mu \PPtot$, 
so putting it all together, the magnetization 
density behaves as
    \beq
        \frac{d \MMtot}{dt} =  n(T) \mudrift Q \BB .
    \eeq
(A stochastic term is omitted).
This apparently explains~\cite{jaubert-mono} the temperature dependences 
of the relaxation times observed years ago by \cite{snyder}.

A slab of ideal spin ice, placed in a magnetic field, responds
(at least in simulations)
just like a slab of conductor placed in an electric field.
After an initial current, surface charge layers form which 
nullify the field in the interior, and the current comes 
to a halt~\cite{jaubert-mono}.

The non-equilibrium recombination dynamics of monopole defects 
with antidefects after a sudden
quench is worked out in Ref.~\cite{castelnovo-quench}, who
suggest its use in real experiments.

\subsection{Wavevector (in)dependent dynamics}
\label{sec:conlon}

Imagine first that the dynamics takes us between states
that all exactly obey the zero-divergence constraint on $\PP(\rr)$.
Then the dynamics of $\PP$ is diffusive and 
Standard dynamics would say
  \begin{eqnarray}
      \frac{\partial \tPP_\perp(\kk)}{\partial t} &=&
      \Gamma(\kk)\frac {\delta \Ftot}{\delta \tPP_\perp(\kk)}  
     + \zeta(\kk,t)\\
      &=& 
       \Gamma(\kk) K \tPP_\perp(\kk)  + \zeta(\kk,t).
  \label{eq:dynamics}
  \end{eqnarray}
I wrote a kinetic coefficient 
``$\Gamma(\kk)$ to allow for different versions of
the relaxational dynamics;
$\zeta(\kk,t)$ is a Gaussian noise source 
with correlation function $2 T \Gamma(\kk)$, which 
satisfies detailed balance, and thus ensures
that the steady state is the Boltzmann distribution
implied by \eqr{eq:Fcoarse}.  
Hence we get a relaxation time
$\tau_0^{-1}(\kk) = \Gamma(\kk) K$.

The naive expectation~\cite{henley-pyrice}
is that, since the dynamics conserves
polarization, $\tau_0^{-1} \propto \Gamma(\kk) \propto |\kk|^2$,
vanishing at small $\kk$.
(That is the location of the static pinch-point singularities;
since the actual variables measured are generally related to $\PP$
by a staggering, they get offset to other wavevectors $\KK$.)
However, for a vector spin model, explicit analytic calculation 
(with a spin-conserving dynamic extension of the large-$n$ approximation
and simulations (with realistic precessional dynamics) 
gave a $\kk$-independent relaxation~\cite{conlon},
i.e. apparently $\Gamma(\kk) \sim$ constant.

This is not the full story; at $T>0$, the flux constraint
is not fully satisfied, so there is a finite correlation length 
$\kappa^{-1}$ (exponential in $1/T$, for a discrete model, or
$\kappa(T)^2 \propto T$ in a vector spin model), as mentioned
above [see Eq.~\eqr{eq:Debye-kappa}].
Ref.~\cite{conlon} found a second 
contribution to the relaxation rate, $\tau_T(T)$,
which represents a current of thermally excited monopoles
(or the analogous excitations in a vector spin model~\cite{conlon}).
The combined formula for the polarization correlations is~\cite{conlon}
  \begin{eqnarray}
      \la P_\mu(\kk,t) P_\nu(-\kk,0)\ra &\propto&
      \Big( \delta_{\mu\nu} - \frac{k_\mu k_\nu}{|\kk|^2} \Big)
                    e^{-t/\tau_T}  \\
         &+&  
      \kappa^2 \frac {k_\mu k_\nu}{|\kk|^2 (|\kk|^2+\kappa^2)} 
              e^{[1/tau_0(\kk)+1/\tau_T] t}.
  \label{eq:P-timecorr}
  \end{eqnarray}
\SAVE{Note that \cite{conlon} considered precessional dynamics of
3-component classical spins, which conserves total spin exactly
at any temperature, as well as polarization only in the $T=0$ 
limit. Is spin conservation the origin of the 
$k_\mu k_\nu$ factor in \eqr{eq:P-timecorr}?}

\section{Quantum models: a quest for ``artificial light''}
\label{sec:quantum}

\SAVE{Could we use the RK mapping to get ``light''
in an explicit, microscopic model?
In particular, can $1/\tau ~ |\qq|$ follow from a model 
with a length-accelerated loop update?  
(Or is there a way to use Chalker's world-line mapping?)}

It is natural to ask, can we have, not just an emergent magnetostatics
or electrostatics, but an emergent {\it electrodynamics} with separate
``electric'' and ``magnetic'' fields satisfying Maxwell's equations?  
Indeed, this can emerge in quantum versions of the models we 
discussed~\cite{hermele,balents-girvin,tewari}.

\OMIT{(Alternate)A quantum mechanical version of the
``Coulomb phase'' naturally has dual electric and magnetic
fields, so 
one might expect ``light'':
excitations with dispersion $\omega\sim |\qq|$ which
are not Goldstone modes \cite{hermele}. }

\subsection{Set-up of quantum models}

Typically, the Hilbert space consists of discrete configurations 
satisfying the same constraints we discussed before, allowing the
definition of a ``magnetic'' field via the same coarse-graining
developed in Sec.~\ref{sec:coarse-grain}.
Furthermore, the model has ``flip''
terms of amplitude $\tring$ which hop you from one basis state to another.  These
are the same kind of ``flips'' (constrained by the flux conservation) 
we found in the classical (Monte Carlo, or physical) dynamics
(see Sec.~\ref{sec:rearrange}
the flip must rearrange a closed loop.  

In a quantum model it is particularly desirable the ``flip''
should look as simple as possible (at least, should be finite!)
when written out in terms of spin or creation/annihilation 
operators.  (Among other things, only the simpler cases are
likely to be realized in experimental systems.)  A ``simple''
flip means a ``minimal update'' which (in the quantum  model)
amounts some variety of ``ring exchange''.  That is, it 
switches the occupancies, or reverses the spins or arrows, along the
shortest loops in the lattice. (These are six sites in the 
cases of the kagom\'e or pyrochlore site lattices).

A natural route to the needed flip terms is to let
the Hilbert space admit local violations 
of the flux constraint, but to make the Hamiltonian strongly penalize
them by a potential energy cost of order $V$. 
The Hamiltonian also includes simpler flips
of amplitude $\tpair$ involving two sites
(spin exchange or boson hopping).
Then a good effective Hamiltonian
(low energy theory) restricts the Hilbert space to the subset of
configurations satisfying the flux constraint, but includes 
loop flip terms generated from perturbation 
theory~\cite{hermele,tewari,banerjee-isakov}: e.g.,
   \beq
       \tring\propto \tpair^3/V^2
   \label{eq:t_ring-tV}
   \eeq
for loops of length six).

\subsection{Theory of the ``Maxwell'' phase}

This theory, written in terms of discrete variables, 
is some kind of ``compact $U(1)$ gauge theory''.
Here $U(1)$ simply means there is one sort of ``electric''
charge. It is the electric flux which has the 
defining constraint I mentioned in Sec.~\ref{sec:intro},
which implies a gauge freedom in the phases of the electric
field operators.~\footnote{
There is not a uniform convention inthe literature whether the variables
are electric and their duals magnetic, or vice versa.}
Also, ``compact'' means the flipping term looks like
$\cos(\Phi_B)$ where $\Phi_B$ plays the role of the 
{\it magnetic} flux, and thus the magnetic flux is
only defined modulo $2\pi$ instead of taking
arbitrarily large values. 
Now, let's assume the $\Phi_B$ fluctuations are rather small.
\footnote{
Since $\EE$ and $\BB$ operators are conjugate, like
position and momentum of a harmonic oscillator, the 
uncertainty relation tells us the ``magnetic'' fluctuations 
can be small only when the ``electric'' fluctuations are large,
as is the case in the sort of correlated-liquid/cooperative-paramagnet
phase we've had in mind throughout this paper.}
Then events (``instantons'') in the quantum evolution in which 
$\Phi_B$ wraps around its period get suppressed, maybe so much 
that we may ignore them, in which case can imagine $\Phi_B$
being non-compact (taking unbounded values) as in the familiar
electrodynamics.  

The next step is usually a leap to a continuum theory, 
by substituting for the ``magnetic'' (flip) term the simplest
analytic (i.e. quadratic) term with the correct symmetries, 
with a coefficient $K_B$ (analogous to the magnetic
permeability); also, a quadratic ``electric'' term is
assumed as in the classical model.
If all that is justified, it follows that we get 
emergent ``artificial light'' elementary excitations~\cite{tewari,hermele}
with dispersion
   \beq
        \hbar \omega(\qq) = \hbar c |\qq|
   \eeq
where $\hbar c \propto \sqrt {K_B/K_E}$ is the analog of the speed
of light.  Although linear dispersing excitations are common
enough in solid state (magnons in antiferromagnets, or phonons),
these ``emergent photons'' are novel in that they are not Goldstone
modes.  I will call such a phase a ``Maxwell'' phase.

The structure factors of such a model do have singularities
at the same pinch-points as the classical models, but the
correlation behavior has different exponents. 
A short cut to the behavior (as in any other model that
reduces to harmonic oscillators) is the virial theorem,
which says the oscillators ``potential'' and ``kinetic'' energy
expectations each get half the zero-point energy.  Thus,
  \beq 
         \la \frac{K_E}{2} |\tEE(\qq)|^2 \ra = 2 (\hbar \omega(\qq)/ 4)
  \eeq
(the 2 is for two polarizations).  The equal-time mode expectations are
thus (in place of the classical result ~\eqr{eq:flucts})
    \beq
           \Struc_E(\qq) \equiv \la {\tEE(\qq)_\alpha}^* {\tEE(\qq)_\beta} \ra \approx
                       \frac {\hbar |\qq| }{\sqrt{K_E K_B}}
                        \Big(\delta_{\alpha\beta}-\khat_\alpha \khat_\beta \Big)
     \label{eq:quantum-mode-corr}
     \eeq
and correspondingly the correlations should decay as $1/r^4$ in the quantum case.
(The structure factor as a function of $(\qq, \omega)$ is similar but
proportional~\cite{tewari} to $|\qq|^2/(\omega^2+ c^2|\qq|^2)$ in place of the $|\qq|$
factor in \eqr{eq:quantum-mode-corr}.)

I somewhat mistrust the leap to the continuum theory, if justified
only by the symmetries of the ``flip'' operator.
Consider the following: given a {\it classical} model with a 
{\it classical} flip dynamics,
one can always devise a quantum Hamiltonian whose
ground state has the same probability distribution as
the (Boltzmann) weights of the classical model,
via the Rokhsar-Kivelson (RK) construction~\cite{RK,henley-RK,claudio-RK}.
But usually the classical system generally
has relaxational dynamics, correspondingly~\cite{henley-RK}
the quantum system has elementary excitations
with $\omega\sim |\qq|^2$ dispersion;
alternatively, if the classical relaxation time is constant
(as in the model summarized in Sec.~\ref{sec:conlon}), 
the quantum  excitations are {\it gapped}.

A second worry is that, as already noted (Sec.~\ref{sec:rearrange}),
a rule that flips only the smallest loops is quite possibly
insufficient to access the whole Hilbert space.

\SAVE{
Another route to a quantum model with manifest electrodynamics
is a classical Coulomb model with 
``accelerated'' (non-local)
updates such that its dynamic exponent is $z=1$, then
use the Rokhsar-Kivelson prescription, so 
the excitations would have linear dispersion.}

\subsection{Simulation tests for emergent electrodynamics}

Two groups have verified the emergent ``Maxwell phase''~\cite{banerjee-isakov,sikora-pollmann}.
Their models are dimer coverings of the diamond lattice (in the case of
\cite{banerjee-isakov}, an equivalent boson model, in which the flip term
is generated as in \eqr{eq:t_ring-tV}.)    The ``smoking gun'' for this state
according to \cite{banerjee-isakov}
is correlations linear in $\qq$ like \eqr{eq:quantum-mode-corr}, while for
\cite{sikora-pollmann} it is an energy depending as $|\EE|^2$  on the uniform flux density.
Unfortunately, with quantum Monte Carlo it would be difficult to measure the 
corresponding dependences for the conjugate variable $\BB$, which ought to 
behave the same as $\EE$.

\section{Phase transitions out of Coulomb phases}
\label{sec:transitions}

\HIDE{Chalker and collaborators studied transitions 
from the Coulomb phase to ordered 
phases~\cite{pickles,powell-dimer-PRL,powell-ice-long,powell-dimer-long}.}

Now that we have (in the polarization field) something 
analogous to an order parameter, we are equipped at 
least to classify transitions out of a Coulomb phase
into, say a long-range ordered phase. We might
even be able to set up a field theory as the basis
for some sort of renormalization group.
But the constraints of the Coulomb phase necessarily
mediate long-range interactions; as is long known~\cite{aharony-dipolar}
these can change the universality class of a transition,
modifying the critical behavior to that characteristic
of a higher spatial dimension.

Since the long-range ordered pattern is a particular
configuration, it has a net polarization $\PP_0$, which can be
used to classify the cases.
\begin{itemize}
\item[]{\it Zero-polarization:}
$\PP_0=0$ is zero;
the transition is a symmetry breaking into one of
several symmetry-related states, all with $\PP_0=0$?  
\item[]{\it Intermediate polarization:}
$\PP_0\neq 0$ so
there are several symmetry-related directions
for $\PP_0$, but each corresponds to just one kind
of ordered state.
\OMIT{The parts of the symmetry
breaking which do and don't change $\PP_0$ are 
qualitatively different, so they should generically 
be broken in two separate transitions.}
\item[]{\it Saturated polarization:}
$\PP_0\neq 0$, with a symmetry breaking as in the
previous case, but with $\PP_0$ at the edge of
the range of possible polarizations, so that
there are no fluctuations whatsoever in the ordered
state (so long as the constraint still holds)
\end{itemize}

Furthermore, such transitions might be driven either by
interactions, or by external fields.

\subsection{Dimer crystal transitions: zero polarization}

The most elementary case is simulated dimer coverings
of a cubic lattice, with an interaction that favors 
dimers on opposite sides of a square plaquette
(in $d=2$, on a square lattice, that would induce the
same columnar pattern found in quantum dimer models.)
The ground state of this Hamiltonian is a periodic pattern 
having $\PP=0$ and with a six-fold degeneracy. (Three directions
the dimers could line up in, and two choices of which layer
to have dimers.)
\SAVE{The natural order parameter corresponds to 
a vector which can point in any of the cubic $[100]$
directions.}

Simulations of this model by Alet {\it et al}\cite{alet-dimers} 
showed a {\it continuous} transition (whereas
Landau theory would predict a first-order transition,
or an intermediate disordered phase).
Furthermore, an emergent $SO(3)$ symmetry was seen in polarization
space (by plotting the distribution of polarizations),
but only right at the transition~\cite{misguich-dimers}.
These properties are suggestive of the recent paradigm of 
``deconfined'' criticality~\cite{deconfined,balents-sachdev} 
whereby some phase transitions 
that must be first order according to Landau's picture, 
instead are continuous, 
and the critical state possesses an emergent symmetry lacking in either of the adjacent phases.  
A flurry (still unresolved) ensued of 
theories~\cite{charrier,powell-dimer-PRL,powell-dimer-long,chen-Q-dimers}
and of followup simulations~\cite{chen-Q-dimers,papanikolaou}

Before these simulations, 
\cite{bergman-PRB} had pondered the transition 
from the Coulomb phase to an ordered 
dimer pattern with zero polarization, concluding it is equivalent to 
the ordering of a superconductor when electromagnetic field fluctuations
are taken into account, which  is supposed to be weakly first order.

Ref.~\cite{chen-Q-dimers} simulated less isotropic versions
of the interaction Hamiltonian. 
The basic picture of the transition is that, coming from the
high-temperature (Coulomb gas) phase, {\it all} of these
cases belong to  the same universality class, an inverted
XY transition.  Those free energy terms anisotropic with respect
to the polarization direction are irrelevant at the critical
fixed point, hence the critical state has the SO(3) symmetry, 
but they are relevant at the attracting fixed point characterizing
the ordered phase.
Thus the transitions are less exotic than in the isotropic case, 
but in one case a surprising phase diagram was found: 
as temperature is raised, the dimer crystal
melted into a paramagnetic phase, then at a higher temperature
entered a Coulomb phase -- re-entrant, in the sense that
it has longer range correlations.
(Notice that even if flux constraints are not enforced in the
microstates, they may appear emergently.)

Ref.~\cite{papanikolaou} elaborated the basic dimer model in
a different way by including further neighbor interactions,
and found a multicritical point (i.e. the transition switches
from continuous to first order) not so far away from
the basic model simulated by \cite{alet-dimers}.

Chalker and Powell~\cite{powell-dimer-PRL,powell-dimer-long}
applied the world-line mapping (see Sec.~\ref{sec:world-lines})
to this problem, and massaged the result into a field theory 
with two complex fields.
The symmetries related to the specific lattice imply that the
terms in this field theory are either $SU(2)$ invariant, or else
of order 8 (hence possibly irrelevant).  This makes plausible
the observed $SO(3)$  (equivalently $SU(2)$) symmetry at the 
transition, which goes with the ``noncompact $CP^1$'' 
(NCCP$^1$) universality class.

\subsection{Intermediate polarization transitions}

Whenever $\PP_0\neq 0$ (and is sufficient to label 
the symmetry-broken states), it should be possible to describe
the transition just in terms of $\PP(\rr)$ and its derivatives.
If $\PP_0$ were not saturated, 
the prescription of Landau theory should lead us more or less
in the right direction: the free energy density $f(\PP)$ 
retains the full symmetry of the lattice,
is analytic in $\PP$, and has global minima 
at the $\PP_0$ values.   Furthermore, rather than 
compute the exact $f(\PP)$, we can usually get away
with Taylor expanding it to the lowest order that has 
minima in the right places.  For example, for 
ordering into a state with $\PP_0$ in one of the six
(100) directions, we might have 
   \beq
    f(\PP) = -\frac{1}{2} \alpha |\PP|^2 
               +\frac{1}{4} \beta |\PP|^4 - 
                   v (P_x^2+ P_y^2 + P_z^2) + \frac{1}{2}\gamma |\nabla \PP|^2
   \label{eq:P-order-100}
   \eeq
When $\alpha, \beta, v \gamma$ are all positive,
the first two terms in \eqr{eq:P-order-100} define a minimum 
along the entire sphere $|\PP|^2= \alpha/\beta$; around that
sphere, the third term is minimized when $\PP$ points in any
of the $(100)$ directions.  The gradient term puts a cost on
the domain wall between ordered domains that have chosen
different ones of the six $\PP_0$ directions.  So  far, this looks
exactly like the Landau-Ginzburg-Wilson ``action'' used
to formulate the $\epsilon$-expansion renormalization
group for a 3-component ferromagnetic spin model
with cubic anisotropy~\cite{aharony-aniso}.
However, we are not done.
If the flux constraint continues to hold through this transition, 
we must enforce $\nabla\cdot\PP =0$ in the 
continuum treatment.
We saw in Sec.~\ref{sec:coarse-grain}  that this constraint generates 
effective dipolar interactions
betwen polarization fluctuations in different places,
and such long-range interactions were shown long ago
to change the critical behavior of spin models.

The case of nonzero, but unsaturated, polarizations 
has the fewest  exotic features.  
It is realized by a Hamiltonian in which
some bond directions are slightly strengthened,
thereby selecting out a unique ground state 
(modulo spin rotation symmetry), so the system undergoes
a N\'eel transition.  
As might be expected, it behaves like the ordering of a 
vector spin system with power-law interactions.
In the vector-spin cases, this is slightly different
from the related models studied in the 1970s~\cite{aharony-dipolar}, 
as the polarization field carries both spatial and spin
component indices: the interaction is anisotropic 
(dipole-like) with respect to the former, but
retains the rotation symmetry of the latter.
\SAVE{The other funny thing that goes on is the
ordering itself suppresses the power-law interaction,
whereas old work treated the latter as a fixed.}

\subsection{Saturated polarization case}

This case has the further complication that,
by the rules of our constraint, no fluctuations are possible 
at all on the ordered side of the transition.  We thus
get the asymmetric behavior called a ``Kasteleyn transition''
\cite{nagle-dimers,jaubert-kasteleyn,powell-ice-long}.
Approaching it from the ordered side, there are no 
critical divergences to signal the impending transition:
in that sense, it looks like  a first order transition.
On the other hand, coming from
the Coulomb phase side, fluctuations are allowed and
in fact have critical divergences as one would expect
from a continuous transition.  Finally, states near  to
the ordered, saturated state can be expressed relative
to it as a set of dilute string excitations (see Sec.~\ref{sec:world-lines})
and that typically implies a relatively tractable behavior.

\subsection{Mapping to $d-1$ quantum problem: World line approach}
\label{sec:world-lines}

Here I summarize a second analytic technique, for Coulomb phases
(beyond plain continuum theory of the polarization field):
 mapping a configuration in $d+1$ 
dimensions to the world-lines of a set of particles in $d$
dimensions. Thus, one is taking advantage of the usual correspondence 
between the partition function of a $(d+1)$-dimensional problem 
in classical stat mech and the path integral of a $d$-dimensional 
quantum many-body problem.  The conservation law of these 
quantum particles corresponds to the defining divergence condition 
of the Coulomb phase model. 

The world-line mapping is ideal in the case of a transitions to 
a phase with saturated polarization,~\cite{powell-ice-long},
since that is where the density of
world-lines goes to zero, but was also applied to a zero-polarization 
case in Refs.~\cite{powell-dimer-PRL,powell-dimer-long}.
It has the disadvantage (compared to the polarization
field theory) that it necessarily breaks the lattice symmetry by
selecting one direction to be the (imaginary) time axis, along which
the world-lines run.   
\SAVE{In general, it has the advantage (over the $\PP(\rr)$ continuum
theory) that it better captures phenomena -- e.g. transitions out 
of the Coulomb phase -- which involve a breakdown of the the polarization 
description is breaking down.}
The set-up is detailed in Appendix~\ref{app:world-line-details}

The Coulomb phase is the disordered phase of the world lines,
which means that to every configuration found in the ensemble,
you can find another configuration in which all the lines 
come out in the same place, except that two of them have got switched.
Exactly as in the ``entangled vortex liquid'' phase, this 
signifies that the bosons in the $2+1$ dimensional quantum model
are in a Bose-condensed phase -- i.e., an ordered (!) phase with
a continuous symmetry-breaking labeled by one angle variable.  
Therefore it has gapless Goldstone modes, the correlations
of which decay as power laws in space and time.  Furthermore,
after one works through the mapping of dimer occupations to
fluxes and of those, in turn, to the boson density and current,
those power laws are precisely the dipolar correlations
expected for the fluxes.

\section{Disorder}
\label{sec:disorder}

Coulomb phase ideas can be applied to disordered cases --
provided that the disorder does not destroy the Coulomb
phase nature of the state.   
If defects are very dilute, the basic phenomenology is that
of the unchanged Coulomb phase between them, as discussed
in earlier sections -- in particular, the nonlocal
spin response in Sec.~\ref{sec:real-space-corr}.
Similarly, disorder might create traps for pseudocharge defects,
(Sec.~\ref{sec:charges}), affecting their phenomenology much as
impurity sites affect carriers in a compensated semiconductor.
Here I address examples of disorder which determine the collective
state of the system.

\subsection{Bond disorder}

One way to gently introduce disorder to a Coulomb phase
is to weakly modulate the interactions in a nearest-neighbor 
antiferromagnet realization (say the pyrochlore lattice), so 
the bond strength has variance $\tDelta^2$~\cite{andreanov}.  
Standard replica techniques gave a mean-field spin glass transition
temperature~\cite{andreanov}
   \beq
           \TcMF = \half \tDelta \Big(\lambda_4 \hatGG \Big)^{1/2}
   \eeq
where $\lambda_4 =6$ is a coefficient derived from the adjacency
matrix, and $\hatGG \equiv\sum _j G_{ij}^2$, where $G_{ij}$ is the
correlation between spins $i$ and $j$.  Where Coulomb-phase
notions enter is just in the evaluation of $\hatGG$.
This has (thanks to our assumption of small $\tDelta$) approximately 
the same numerical value as the pure case, for which we know 
the asymptotic behavior (Sec.~\ref{sec:correlations}), and indeed
(within the large-$N$ approximation) the exact function~\cite{garanin-canals}.

Alternatively, one considers {\it strong} random modulation
but in dilute places.  This produces local pseudospin degrees
of freedom, which have effective long-range interactions
mediated by the polarization field,~\cite{andreanov} as we knew from 
Sec.~\ref{sec:coarse-grain} and Sec.~\ref{sec:charges}.~\footnote{
This is reminscent of Villain's pseudospins induced at wrong bonds,
which had pseudo-dipolar interactions mediated by the Goldstone
mode of an ordered antiferromagnet~\cite{villain79}.
}
Thus the problem maps to a dipolar magnet dilutely occupied by spins,
also giving a spin glass phase, as known for several decades.

\subsection{Depleted antiferromagnets}
\label{sec:depleted}

The ``hyperkagom\'e'' antiferromagnetic lattice~\cite{Ok07,Ho07}
is a pyrochlore lattice with 1/4 of 
the magnetic sites removed, making very regular pattern of 
corner-sharing triangles. 
\SAVE{With nearest-neighbor coupling it is equivalent to the half-garnet lattice,
see Sec.~\ref{sec:examples}.}
What if we dilute at random, but place the non-magnetic 
substitutions such that exactly one site gets removed from every 
tetrahedron~\cite{henley-depleted}.
This ensemble is equivalent to the dimer coverings of the diamond lattice
(where each dimer corresponds to a removed site), and thus is highly
plausible as an example of a lattice-gas realization~(see Sec.~\ref{sec:examples}).
The ``flux'' can still be defined, and is still conserved, on
such ``randomly depleted lattices''.  Hence, they still
have Coulomb phases despite the disorder~\cite{henley-depleted}.
The disorder acts on the coarse-grained polarization field 
like a random anisotropy term with long-range dipolar-like correlations,
since the randomness itself is a Coulomb phase ensemble~\cite{henley-depleted}.

\begin{figure}
\centerline{
\includegraphics[width=11.7cm,angle=0] {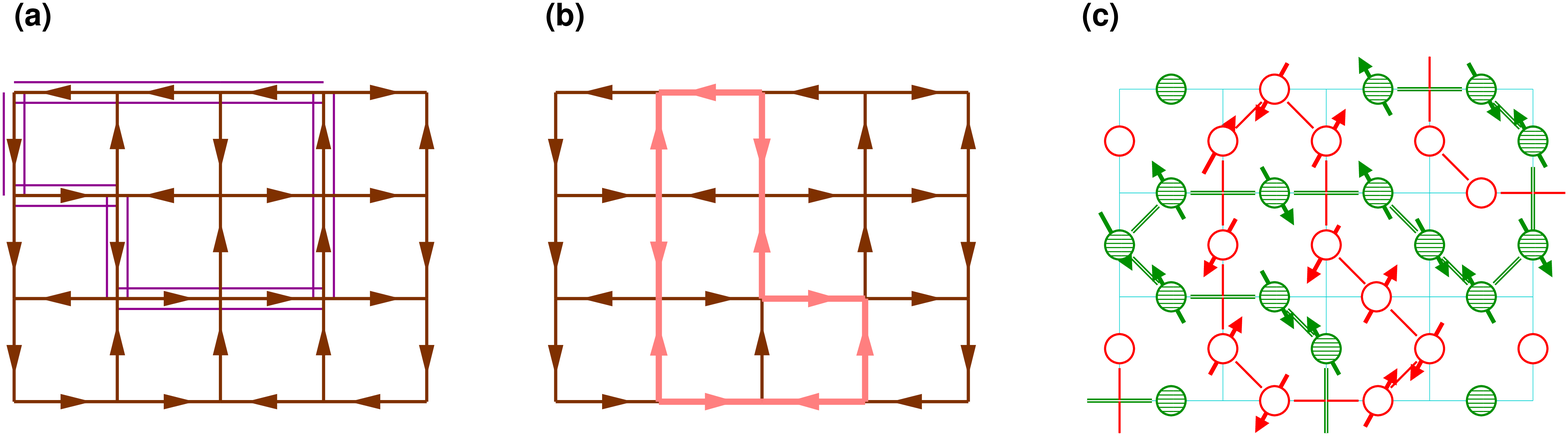}
}
\caption{Loops in ice model and constrained disorder.
(a). An ice-arrow configuration; outlined is an update
loop (reversing all arrows gives another valid configuration).
(b). The same configuration; highlighted (pink arrows)
is an alternating loop  (ice-arrows point in alternating
directions)
(c). A lattice gas of A and B atoms (red and green circles)
placed so that every vertex is surrounded
by exactly two of each species, and represented by the same
ice-arrow configuration as (a). 
(Each red atom maps to an arrow from an even to the odd
vertex, and oppositely for each green atom.)
Loops connecting atoms of the same species
form alternating loop.
Banks~\cite{banks} showed that for some plausible
species-dependent couplings (in the three-dimensional analog),
the spin directions alternate along each loop as shown.}
\label{fig:loops-all}
\end{figure}

\subsection{Constrained disorder models: loop-based dilution}
\label{sec:constrained-disorder}

Besides the depleted lattices,
there are other examples of the (nearly unstudied)
class of correlated-disorder ensembles that I'll call ``constrained disorder'',
meaning the frozen configurations come from some ``Coulomb phase'' ensemble.
For example, 
it was proposed by \cite{banks} that in CsNiCrF$_6$ \cite{zinkin97}, 
A (=Cr) and B (=Ni) spins populate the pyrochlore such that  every tetrahedron 
has two of each.  
As noted in Sec.~\ref{sec:examples}, this realizes an ice model.

The spin-spin exchange constants are
$J_{AA}$, $J_{AB}$, and $J_{BB}$, depending on the species
at either end of a bond.  In a realistic range of couplings,
the $A$ spins in each tetrahedron point along $+\nn$ and $-\nn$ while the $B$
spins point along an independent direction $+\nn'$ and $-\nn'$.
Hence, the lattice breaks up into disjoint loops of $A$ or $B$ spins;
in each loop spin directions alternate, and 
each loop chooses a staggered direction 
independent of the others [Fig.~\ref{fig:loops-all}(c)]
Thus the disorder-averaged spin correlation $C_{ij}$ is simply 
the connectedness correlation function -- the probability 
that sites $i$ and $j$ are on the same loop. 
In simulations \cite{banks}, $C_{ij} \propto 1/r_{ij}^a$
with $a\cong 1$.

Consider flux-loops, as mentioned in Sec.~\ref{sec:rearrange},
and also the basis of world-lines in Sec.~\ref{sec:world-lines}:
flux runs in the same sense along the loop or line.
One can make a one-to-one correspondence between configurations
which have a flux-loop containing both sites $\rr$ and $\rr'$, 
to configurations with a charge defect at $\rr$ and an antidefect at 
$\rr'$.  (Namely, one maps the former to the latter by reversing the flux
along half the loop between $\rr$ and $\rr'$. On the other hand,
if one has a charge defect, it must be the endpoint of two such flux 
lines; and if the system has only one defect pair, the other endpoints
must be at the second defect, so the map is invertible.)
This idea -- originating with \cite{saleur-duplantier} -- was
applied in two-dimensional height models to get the connectedness correlation
function~\cite{kondev-4color,kondev,kondev-loops}, 
and the same idea will work in Coulomb phases.
However, the loops in question here are of a different kind:
they run alternately with and against the flux, they cannot be reversed,
so the defect-pair trick is not applicable.  It will be interesting
to see how a simple power-law can emerge for the connectedness correlation
function.
The same kind of loops are formed by fermions in the wavefunctions
envisaged by Fulde {\it et al}~\cite{fulde-LiV2O4,fulde}.

\section{Conclusion}

When local constraints imply the existence of a conserved ``flux'' 
in some statistical model, this flux can be
coarse-grained into a polarization field which is a comprehensive
framework -- not yet exhausted -- to map out the model's behaviors:
long-range correlations, charge-like defects, dynamics, and disorder
effects. 

In the course of the review, I've outlined the associated
calculational tools:
\begin{itemize}
\item[(1)] Continuum theories, e.g. to model the analytic form of
correlations in classical or quantum models;
\item[(2)] The large-$n$ limit of $n$-component spin models, to evaluate
correlations in particular lattices.
\item[(3)] ``World lines'' (or equivalently ``Dirac strings'') near the
limit of saturated polarization, e.g. the behavior of ``spin ice''
magnets in large fields of the appropriate orientations.
\end{itemize}

\ack
I acknowledge helpful conversations and communications with
Kedar Damle, Arnab Sen, Claudio Castelnovo, Roderich Moessner, 
Simon Banks, Nic Shannon, Frank Pollmann, Simon Trebst,
and especially John Chalker.
This was supported by the NSF under grant DMR-0552461.
I am grateful to Boston University for hospitality
during the writing of this paper.

\appendix
\section{Set-up of world-line mapping}
\label{app:world-line-details}

Here I present details of the world-line approach, applied to a variety 
of questions about Coulomb phases by Chalker and collaborators.
THe key idea is to map a configuration in $d+1$ 
dimensions to the world-lines of a set of particles in $d$
dimensions. Thus, one is taking advantage of the usual correspondence 
between the partition function of a $(d+1)$-dimensional problem 

For pedagogical purposes, my illustrations are from the $1+1$ 
dimensional cases, although (as noted already in Sec.~\ref{sec:height})
these are ``height'' models and thus have much more structure than the 
$2+1$ dimensional Coulomb phase models.
Specifically in the world-line mapping, additional features in
$1+1$ dimension are (i). if the model is exactly soluble, this is how one 
sets it up (e.g. for the Bethe ansatz)  (ii). usually the world lines
are not crossing, so they represent hard core bosons; in one dimension,
one can re-interpret the particles (via a Jordan-Wigner transformation)
as fermions, and the statistics naturally implements the hard core
constraint.  In $2+1$ dimensions, one is stuck with hardcore 
(thus, interacting) bosons.  Since one is mainly after long-wavelength
properties, the repulsive interactions are not a big concern:  
one coarse-grains beyond the scale of the boson spacing, and
describes the boson fluid using just (quantum) hydrodynamics.

Our mapping must ensure that the world lines don't turn back on themselves
(so that the particles are conserved and don't annhilate).   This is done
by letting the world lines be the difference between the actual 
configuration and a reference configuration
having saturated polarization in some reference direction. 
If the original configuration had a nonzero mean flux
along the reference direction, the resulting array of 
world lines has a different density from the reference;
while if the original flux was transverse, the resulting
array becomes tilted, representing a nonzero current 
in the lower dimensional quantum system.

\begin{figure}
\centerline{\includegraphics[width=11.0cm,angle=0] {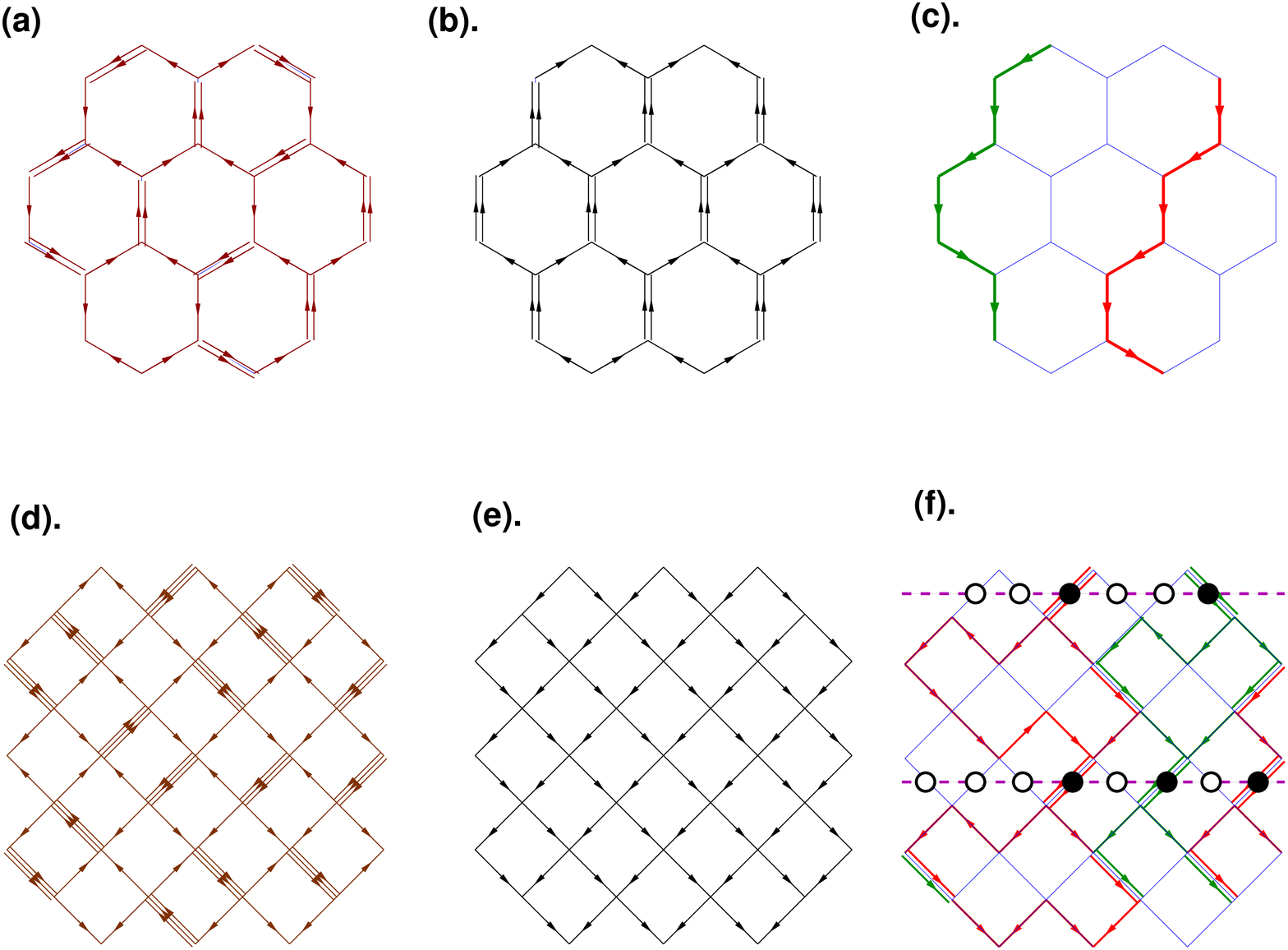}}
\caption{
Mapping configurations of a flux model to world lines.
Parts (a,b,c) illustrate the dimer-difference construction.
(a) A typical dimer covering on the honeycomb lattice,
with arrows giving the fluxes; single arrows are 
uncovered edges.  (b) A reference dimer covering, having
the maximum possible flux (in the $=y$ direction).
(c).  The difference between the fluxes from (a) and (b) 
forms an array of non-intersecting lines that always 
move from top to bottom, and may thus be interpreted
as world lines of a set of particles.
Parts (d,e,f) show the uniform-flux construction on
the square lattice: again, (d) is a typical dimer
covering, (e) is the reference flux pattern [but
unlike (b), it does not represent any dimer covering], 
and  (f) is the difference, showing world lines.
When we slice through every fourth layer of bonds
[dotted lines in (f)], the lines are always paired
in that layer, and we interpret every pair as
a particle [filled circle] in a chain of sites.
(In (c) and (f), all world lines are equivalent, but
those which go with different particles are 
alternately colored red and green.)
}
\label{fig:WL}
\end{figure}

Figure~\ref{fig:WL} illustrates two classes of this
construction.  Fig.~\ref{fig:WL}(a,b,c) is the
``dimer-difference'' construction.  Quite generally,
if you take two dimer coverings of the same lattice
and fill in edges wherever they differ, you get a set
of unbroken lines or loops. If one of the coverings 
had saturated flux in a certain direction, these lines 
are directed along that same direction. This construction
was used in \cite{adhar}. 

\SAVE{As an aside, it represents the simplest path to an 
Onsager-type exact solution of any model -- possible,
of course, only in two dimensions.
Just interpret the world-lines as independent fermions.
It turns out that Fermi statistics simply cancel
the weights of configurations in which world-lines would 
have touched, but do not introduce negative weights.
One just needs to compute the transfer-matrix eigenvalues
$\lambda(k_x)$ for the wandering of a single world line, 
which is easy if one is not bothered by the offsetting of 
the lattice of sites between one layer and the next one.  
Then $\ln \lambda(k_x)$ is a sort of dispersion relation;
one just fills a Fermi sea and
$-\int _{-k_F}^{+k_F} dk_x \ln \lambda(k_x)$ gives the
actual entropy density of random dimer coverings.}

I call the other construction a ``uniform-flux 
construction'';  this is implicit in the set-up
used by Powell and Chalker~\cite{powell-dimer-PRL,powell-dimer-long}.
They define it for dimers on a simple-cubic lattice, which 
is oriented with the reference being
a (111) axis; Fig.~\ref{fig:WL}(d,e,f) shows the square lattice analog.
The diagonal orientation of the world lines has the
virtue that coordinate-axis directions remain symmetry-equivalent;
(Of course, one cannot manage that on other lattices, e.g. the honeycomb;
and if you wished the (11) type axes to remain equivalent, you would 
pick the other orientation.)
In the uniform-flux construction, the flux lines are not literally world-lines,
since they can turn momenentarily backwards in the odd layers; furthermore
each ``particle'' has two flux lines which split apart.  However, if
one looks at the particles only in every fourth layer (dashed slices in
Fig.~\ref{fig:WL}(d), the lines always move forwards in time (downwards
in the figure) and always rejoin in the same pairs.  In the simple
cubic lattice, the slices should be taken every six layers, and each
particle splits into three flux lines, which won't always rejoin
but usually do.  Rather than compute the exact transfer matrix
eigenvalue $\lambda^6(q)$, Powell and Chalker just use a function
with the same behavior near its maximum, since they are only
interested in long-wavelength behaviors (and will get those by 
mapping the $d=2+1$ problem to a field theory).

\end{document}